\documentclass[twocolumn]{svjour3}           

\smartqed  

\usepackage{graphicx}
\usepackage{amssymb}
\usepackage{amsmath}
\usepackage{nicefrac}

\journalname{Tribology Letters}

\begin{document}

\title{
On the contact area and mean gap of rough, elastic contacts
}
\subtitle{Dimensional analysis, numerical corrections and reference data}


\author{Nikolay Prodanov \and
        Wolf B. Dapp \and
        Martin H. M\"user
}


\institute{Nikolay Prodanov \and
  	       Wolf B. Dapp \and
           Martin H. M\"user \at
           J\"ulich Supercomputing Centre \\
           Institute for Advanced Simulation \\
           FZ J\"ulich, 52425 J\"ulich, Germany\\
           \and
           Nikolay Prodanov \and
           Martin H. M\"user \at
           Department of Materials Science and Engineering\\
           Saarland University, Campus  \\
           66123 Saarbr\"ucken, Germany \\
           \email{martin.mueser@mx.uni-saarland.de}
}

\date{Received: date / Accepted: date}

\maketitle

\begin{abstract} 
The description of elastic, nonadhesive contacts between solids with
self-affine surface roughness seems to necessitate knowledge of a
large number of parameters.
However, few parameters suffice to determine many important interfacial
properties as we show by combining dimensional analysis with
numerical simulations.
This insight is used to deduce the pressure dependence of the relative
contact area and the mean interfacial separation $\Delta \bar{u}$ and to
present the results in a compact form.
Given a proper unit choice for pressure $p$, i.e.,
effective modulus $E^*$ times the root-mean-square gradient $\bar{g}$,
the relative contact area mainly depends on $p$ but barely
on the Hurst exponent $H$ even at large $p$.
When using the root-mean-square height $\bar{h}$ as unit of length,
$\Delta \bar{u}$ additionally depends
on the ratio of the height spectrum cutoffs at short and long wavelengths.
In the fractal limit, where that ratio is zero, solely the roughness
at short wavelengths is relevant for $\Delta \bar{u}$.
This limit, however, should not be relevant for practical applications.
Our work contains a brief summary of the employed numerical method
Green's function molecular dynamics including an illustration of how to
systematically overcome numerical shortcomings through appropriate
finite-size, fractal, and discretization corrections.
Additionally, we outline the derivation of Persson theory in dimensionless
units.
Persson theory compares well to the numerical reference data.
\end{abstract}

\section{Introduction}

Most solids have surfaces with self-affine roughness, which means that
the height spectra of their undeformed surfaces scale with a power of the wave vector over several decades~\cite{Power91,Persson05JPCM,Lechenault10}.
As a consequence of this roughness, solids tend to touch intimately only
at a miniscule fraction of the apparent contact area~\cite{Prodanov13}.
Central quantities characterizing mechanical contact are the
relative contact area $a_{\rm r}$, the mean gap $\Delta \bar{u}$
between the solids, and the contact stiffness $K$, which is the derivative
of $\Delta \bar{u}$ with respect to pressure~\cite{Lyashenko13,Pastewka13,Barber13}.
Predicting either one of those descriptors for a given system --- with
well-defined height spectra and elastic properties ---
had not been possible until the beginning of the last decade.
This changed when Persson proposed a scaling approach to contact
mechanics and rubber friction~\cite{Persson01}. The theory prompted the first numerical simulations which
addressed systematically the contact mechanics of solids with self-affine
rough surfaces~\cite{Hyun04}.

Traditional approaches to contact mechanics neglect long-range
elasticity~\cite{Greenwood66,Bush75,Persson06}.
This approximation is not only undesirable from a mathematical point of view,
because it is uncontrolled, but also for practical reasons.
Calculations neglecting long-range elasticity
almost always lead to qualitatively wrong results~\cite{Lyashenko13,Persson06}.
One example is that traditional contact theories predict that
the gap distribution remains Gaussian even under load when in fact it is exponential.
As a consequence, traditional approaches to contacts
grossly underestimate
by how much additional load reduces the mean gap.
This can easily lead to a several-decades overestimation of the leakage current
through a mechanical interface at a given relative contact area even far
away from the percolation transition~\cite{Dapp12PRL}.

In contrast, Persson theory has passed all comparisons to simulations
so far, in the sense that the correct functional dependencies or
constitutive laws follow from
it, at least for relative contact areas of less than 50\%.
The coefficients of the constitutive laws tend to be within ${\cal O}(10\%)$
of those produced by high-quality simulations~\cite{Pastewka13,Dapp12PRL,Almqvist11,Campana07,Putignano12}.
For example, it predicts the area-load dependence and gap distribution
functions to within 15\% accuracy between extremely small and
50\% contact area~\cite{Almqvist11,Campana07,Putignano12}.

The field of contact mechanics of randomly rough surfaces has much matured in the
last ten years.
Persson extended his theory to many interfacial problems,
such as adhesion~\cite{Kendall01,Persson01adhes,Lorenz13jpcm}, plasticity~\cite{Persson06,Aifantis87}, contact stiffness~\cite{Pastewka13,Campana11}, leakage~\cite{Persson08JPCM,Lorenz10EPJE,Persson12EPJE}, squeeze-out~\cite{Lorenz10EPJE2},
and the transition from elastohydrodynamic to boundary lubrication~\cite{Persson09JPCM}.
As far as numerics are concerned,
various groups now have the expertise to simulate solids with surfaces
containing several million grid points in the top layer,
although considering billions of grid points per plane
(as we did in the preparation of this work) still necessitates
the use of supercomputers.
Despite much progress, well-defined reference data is lacking
and, moreover, estimates for the ratio of relative contact area and dimensionless load
at small load have not yet
converged~\cite{Hyun04,Campana07,Putignano12,Yastrebov12}.
One of the reasons for the lack of such data is that no
dimensional analysis has been conducted assessing which parameters are relevant and which quantities should be used to dimensionalize data.
Such an analysis is useful to simplify the comparison between
experiment, theory, and simulation.

In this work, we provide reference data for two interfacial properties:
$a_{\rm r}$ and $\Delta \bar{u}$, 
including instructions on how to dimensionalize the data. Note that the contact stiffness $K$ can subsequently be obtained from the mean gap, which is the reason why we do not
consider $K$ in the present work.
One aim of our endeavor is to put experimentalists in a position to
deduce reasonable approximations to, say, the mean gap
as a function of pressure.
At the same time, we wish to enable theoretically inclined practitioners to deduce some of the answers by themselves. For this purpose, we review Persson's contact mechanics theory
(finding occasional shortcuts to some of the original calculations)
and also describe the Green's function molecular dynamics (GFMD) method~\cite{Campana06,Kong09},
which allows one to conduct simulations of planes with several
million grid points on a standard single CPU core.
This includes guidelines on how to systematically extrapolate the
observable of interest to large system sizes (finite-size scaling),
to large ratios of short- and long-wavelength cutoffs (fractal scaling),
and to the continuum limit.
Last but not least, we introduce a way to nondimensionalize the
data with the goal to facilitate the comparison of data from
different research groups.

The paper is organized as follows. In Section~\ref{sec:theory} we introduce the basic model assumptions and describe the means of characterizing the surface roughness.
Section~\ref{sec:results} presents the results of GFMD calculations for the scaling of relative contact area and mean gap with different corrections. Section~\ref{sec:results} also contains the reference data. Section~\ref{sec:conclusions} summarizes the main findings.
In the appendix, we briefly review Persson's theory as well as the GFMD technique. 

\section{Theory and method}
\label{sec:theory}

\subsection{Basic model assumptions}

Throughout this work we make some basic assumptions and approximations.
These are (i) linear elasticity of the solids, (ii) hard-wall
repulsion between them, and (iii) the small-slope approximation.
We make no additional (uncontrolled) approximations  in our numerical
solutions of the contact problem, unlike traditional contact mechanics
approaches such as the Greenwood-Williamson theory~\cite{Greenwood66,Persson06}, one assumption of which are circular or elliptical shapes
of contact patches (numerical simulations reveal that contact predominantly lives
in fractal patches, which arise through the merging of many Hertzian contacts).
We note that dropping any of our assumptions would make
it impossible to present a complete set of reference data without making
use of three-dimensional representations or tables. This means that plastic deformation and adhesion of the surfaces are not included in our model.

Nevertheless, our assumptions can be considered to be a good approximation for many applications. Plasticity only becomes relevant when the locally averaged stress
in a contact exceeds a threshold, namely the hardness.
At macroscopic scales, the stress is much smaller than the macroscopic
hardness.
When fine features of the height or the contact geometry are resolved,
stresses become large, but so does the hardness, which is a
scale-dependent quantity~\cite{Ma12}.
Persson theory allows one to estimate these effects and reveals that
for many quantities plasticity will only induce small perturbations.

Likewise, direct adhesive interactions between solids are confined to
those points where the two surfaces touch microscopically or are
about to do so. Nevertheless, adhesion can become relevant (for soft solids) when capillaries
are present.
We point the reader to Refs.~\cite{Persson01adhes} and~\cite{Lorenz13jpcm}
for a more detailed discussion.
In this work we assume, as already mentioned, hard-wall interactions,
i.e., at no position at the interface may the $z$-coordinate
of the top solid be smaller than that of the substrate.
Formally, this nonholonomic boundary condition can be written as:
\begin{equation}
z_{\rm top}(x,y) \ge z_{\rm bottom}(x,y),
\label{eq:boundaryCond}
\end{equation}
where $z_{\rm top}(x,y)$ and $z_{\rm bottom}(x,y)$ are the $z$-coordinates
of the contacting surfaces for top and bottom solid, respectively,
while $x$ and $y$ indicate in-plane coordinates.
No forces act between the solids when they are not in contact and infinitely
high repulsion occurs when they overlap.

The assumption of linear elasticity together with the small-slope approximation
makes it possible to combine both roughness and compliance of the two solids
into an effective roughness and an effective compliance~\cite{Johnson85}.
These can then be assigned to either side
of the interface.
The effective local height then becomes
$h(x,y) = z_{\rm top}(x,y) - z_{\rm bottom}(x,y)$ and
the effective modulus reads
\begin{equation}
\frac{1}{E^*} =
\frac{1-\nu_1^2}{E_1} + \frac{1-\nu_2^2}{E_2},
\label{eq:effMod}
\end{equation}
where the $E_i$ and $\nu_i$ denote the elastic moduli and the Poisson
ratios of the two contacting solids.

The small-slope approximation permits neglecting the sideways
motion of atoms, so that it suffices to consider scalar displacement
fields.
Put differently, by restricting ourselves to a scalar displacement field
and by using the effective modulus $E^*$, we implicitly implement
the small-slope approximation, even if the slopes for which we solve
the contact mechanics problem are large.

\subsection{Quantifying surface roughness}

Let us define the effective height of the undeformed interface $h(x,y)$
as the gap between the surfaces when they touch in a single
point, i.e., in the limit of a vanishingly small normal load $L = 0^+$.
The values of $h(x,y)$ can be interpreted as a field of (spatially correlated)
random numbers.
The main assumption usually made for their  stochastic properties
is homogeneity across the surface, e.g., no wear tracks or systematic
surface structuring.
In addition, we will also assume isotropy, i.e., a Peklenik number of one.
As a consequence of homogeneity and isotropy any point is expected to yield
the same average height for different realizations of the interface.
Moreover, local gradients average to zero, although their magnitudes
are finite.
Lastly, the expected magnitude of the height change between two points
(on a self-affine surfaces)
increases as a power law of the distance between them.
This can be expressed mathematically as:
\begin{eqnarray}
\langle h({\bf r}) \rangle & = & \bar{h}_0, \\
\langle h({\bf r}) - h({\bf r}+ \Delta {\bf r}) \rangle & = & 0, \\
\left\langle
  \left\{ h({\bf r}) - h({\bf r}+\Delta {\bf r}) \right\}^2
\right\rangle
& \propto & \Delta r ^ {2H},
\end{eqnarray}
where ${\bf r}$ and $\Delta {\bf r}$ are vectors in the $(x,y)$ plane.
$H$ is the Hurst roughness exponent~\cite{Persson05JPCM}, which also determines
the fractal dimension of a surface, i.e.,  $D_{\rm f} = 3-H$.

In Fourier space, the stochastic properties of the surface roughness read:
\begin{eqnarray}
\left\langle \tilde{h}({\bf q}) \right\rangle & = & 0
\mbox{ for } q\ne 0,  \\
 \left\langle \tilde{h}^*({\bf q}) \tilde{h}({\bf q}') \right\rangle
 & = &
\delta_{{\bf q},{\bf q}'} \, C({\bf q}),
\label{eq:WienerKhinshin}
\end{eqnarray}
where the surface height spectrum $C({\bf q})$ exhibits the power law
scaling
\begin{equation}
\label{eq:scalCinRec}
C({\bf q}) = C(q_0) \left(\frac{q}{q_0} \right)^{-2-2H}
\end{equation}
within a range
$2\pi/\lambda_{\rm l} < q < 2\pi/\lambda_{\rm s}$, i.e., in between
cutoffs at long and at short wavelengths, respectively.
$q_0$ indicates an arbitrary reference wavenumber, which one can
choose to coincide with $q_l = 2\pi/\lambda_{\rm l}$.
In Eq.~(\ref{eq:WienerKhinshin}), $\delta_{{\bf q},{\bf q}'}$
represents the Kronecker symbol, which needs to be replaced with the
$\delta({\bf q}-{\bf q}')$
Dirac delta function for infinite systems in the continuum limit instead of discrete, periodically
repeated systems.

For experimental systems~\cite{Persson05JPCM} $0\le H \le 1$. While it is formally possible to assume values outside this interval, we are not aware of any experiment finding such Hurst exponents, so we disregard that possibility here.
The typical situation is that a surface power spectrum has a rolloff~\cite{Persson05JPCM} at wave vector
$q_{\rm r}$ so that $H=-1$ for $q<q_{\rm r}$ is a reasonable approximation
and $H=0.85 \pm 0.05$ for $q>q_{\rm r}$.
From a computational point of view, a rolloff is as easily implemented
as in a theoretical approach.
We nevertheless disregard the rolloff here and instead focus on the
contact mechanics for wavelengths shorter than the rolloff wavelength.
Our motivation for this choice is that this makes a comparison between
theory and simulation more transparent.
Moreover, it is not possible to produce meaningful reference data
when the limits ${\lambda_{\rm r}/{\cal L}}\to 0$ and
${\lambda_{\rm s}/{\lambda_{\rm r}}}\to 0$ do not interchange,
where ${\cal L}$ is the linear system size.
When using a long wavelength cutoff rather than a rolloff,
the interchangeability of limits is much less problematic.

The stochastic properties of the surface --- or an interface --- are
fully defined by the following variables:
$H$, ${\cal L}$, $\lambda_{\rm l}$, $\lambda_{\rm s}$, $C(q_{\rm l})$,
and, in the case of numerical calculations, $a$, which is the resolution
of the lattice, i.e., the (smallest) grid spacing of the discrete
elastic manifold.
These six parameters can be replaced by the following set of parameters:
$H$, $\bar{h}$, $\bar{g}$,
${\cal L}' = {\cal L}/ \lambda_{\rm l}$,
$\lambda'_{\rm l} = \lambda_{\rm l} / \lambda_{\rm s}$, and
$\lambda'_{\rm s} = \lambda_{\rm s} / a$.

The RMS height $\bar{h}$ and RMS gradient $\bar{g}$ can be computed
either in real space or in Fourier space. For a discrete set of
heights we use the transform
\begin{eqnarray}
h({\bf r}) & = & \sum_{\bf q} \tilde{h}({\bf q}) \exp[i {\bf q}\cdot {\bf r}] \\
\tilde{h}({\bf q}) & = &
\frac{1}{N} \sum_{\bf r} \tilde{h}({\bf r}) \exp[-i {\bf q}\cdot {\bf r}],
\end{eqnarray}
where $N$ is the number of points in the surface, and where ${\bf q}$ should
be chosen in the image in which $q$ is minimized.
Thus,
\begin{eqnarray}
\bar{h}^2 & = & \frac{1}{N} \sum_{\bf q} \vert \tilde{h}({\bf q}) \vert^2,
\label{eq:barH} \\
\bar{g}^2 & = & \frac{1}{N} \sum_{\bf q} q^2 \vert \tilde{h}({\bf q})
\vert^{2},
\label{eq:barG}
\end{eqnarray}
for discrete systems.
In order to connect to continuum theories, one needs to replace the discrete Fourier sums for finite, discrete and periodically repeated systems
with Fourier integrals representing infinite and continuous systems. Going from the discrete to continuous representation in Fourier space implies that the thermodynamic limit ${\cal L}\rightarrow\infty$ is satisfied. Taking into account that nominal contact area $A_0\rightarrow\infty$ when ${\cal L}\rightarrow\infty$, the following expression should be used:
\begin{eqnarray}
\sum_{\bf q} & \to & \lim_{A_0\rightarrow\infty}\frac{A_0}{(2\pi)^2} \int d^2q.
\end{eqnarray}
Evaluating $\bar{h}$ and $\bar{g}$ for continuous systems analytically (for $0 < H < 1$),
one can recognize that they are dominated by long and short wavelengths, respectively:
\begin{eqnarray}
\label{eq:scalHeight}
\bar{h}^2   & = & \frac{q_{\rm l}^2 C(q_{\rm l})}{H}
                  \left\{1- \left(q_{\rm l}/q_{\rm s}\right)^{2H}\right\}, \\
\label{eq:scalGradient}
\bar{g}^2 & = & \frac{q_{\rm s}^4 C(q_{\rm s})}{2-2H}
 \left\{ 1-\left(q_{\rm l}/q_{\rm s}\right)^{2-2H}
 \right\},
\end{eqnarray}
because $\epsilon_{\rm f} \equiv q_{\rm l}/q_{\rm s}$
disappears in the ``fractal'' limit (defined as $\epsilon_{\rm f}\to 0$).
Specifically, $\bar{h}_{\rm f} = q_{\rm l} \sqrt{C(q_{\rm l})/H}$
(for $H>0$) depends only on the spectral features at $q_{\rm l}$ in the
fractal limit, while $\bar{g}_{\rm f} = q_{\rm s}^2 \sqrt{C(q_{\rm s})/(2-2H)}$
(for $H<1$)
only depends on the spectral features near $q_{\rm s}$.

Any observable ${\cal O}$ measured or computed for a given normal pressure $p$
will be a function of many variables, i.e.,
\begin{equation}
{\cal O} = {\cal O}(p,E^*,H,\bar{h},\bar{g},1/{\cal L}',1/\lambda'_{\rm l},
1/\lambda'_{\rm s}).
\end{equation}
However, in most cases of practical interest, one should be close to
the following two limits:
(i) the thermodynamic limit $1/{\cal L}'\to 0$ and
(ii) the fractal limit $1/\lambda'_{\rm l} \equiv \lambda_{\rm s}/\lambda_{\rm l} \to 0$.
Moreover, when comparing to continuum theories, one should reduce
discretization effects, which leads to
(iii) the continuum limit $1/\lambda'_{\rm s} \to 0$.
In most cases, one should therefore be interested only in the dependence
of a quantity on three surface-topography-related variables, namely
$H$, $\bar{h}$, and $\bar{g}$.
Since we are still free to choose the unit of length, $H$ and $\bar{g}$
are the only dimensionless parameters that can matter
in the thermodynamic/fractal/continuum (TFC) limit.

In consequence,
the mean gap in a  self-affine fractal interface in the TFC limit
can only depend on three variables, i.e.,
\begin{equation}
\Delta \bar{u} = \bar{h} \cdot \Delta u_{\rm dl}(p/E^*,H,\bar{g}),
\label{eq:meanGapStructure}
\end{equation}
where $\Delta u_{\rm dl}$ is the dimensionless mean gap, which can
only depend on dimensionless numbers.
Any other dependence is not possible, because $\Delta \bar{u}$ must have
the dimension of length.
Of course, Eq.~(\ref{eq:meanGapStructure}) is only meaningful
under the assumption that the TFC limit exists and is unique, i.e., that
it does not matter in what order the limits
${1/{\cal L}'} \to 0$ and ${1/{\lambda}_{\rm l}'} \to 0$
are taken.

Given the above analysis, one can conclude that not only the gap but
\textit{any} quantity can depend in a nontrivial fashion on at most two surface
topography-related, dimensionless parameters, namely $H$ and $\bar{g}$.
This constitutes a dramatic reduction of complexity as compared to the
initial set of six topography-defining variables.
As a caveat we note that it is nevertheless possible that
$\Delta u_{\rm dl} = 0$. In that case,
the leading order is a correction $\propto \epsilon _{\rm f}$.

\subsection{Dimensional analysis of the elasticity of half spaces}
\label{sec:dimAnal}

In three-dimensional space, the elastic energy density $U_{\rm el}/V$
is a bilinear function of the strain
$(\partial u_\alpha / \partial R_\beta + \partial u_\beta/\partial R_\alpha)/2$
in the harmonic approximation. Here, $u_\alpha({\bf R})$ is the displacement field,
and it is proportional to the elements
of the elastic tensor $C_{\alpha\beta\gamma\delta}$~\cite{Landau70v7}.
For a homogeneous medium this implies that the energy density is bilinear in
the wave vector if the energy is calculated in Fourier space.
The elasticity of a half-space must still be harmonic in the displacements
(which are now only defined on the surface) and it will still be proportional
to the elements of the elastic tensor.
However, the energy density is no longer normalized to a volume element but
to a surface element.
Since the areal energy density must still be harmonic and thus quadratic
in the displacements, the prefactor can only be proportional to the wave vector. This means that
$U \propto q E^*\vert \tilde{u}({\bf q}) \vert^2$ is the only
possible dependence.
Fixing the prefactor requires lengthy calculations~\cite{Persson01adhes}, the result of
which is
\begin{equation}
U_{\rm el}/A_0 = \frac{E^*}{4} \sum_{\bf q} q \vert \tilde{u}({\bf q}) \vert^2,
\label{eq:elaEnerg2D}
\end{equation}
where  $A_0$ is the nominal surface area of the solid experiencing an
external force.
In Eq.~(\ref{eq:elaEnerg2D}), we have restricted the displacement to be
normal to the surface, which is justified in the small-slope approximation
as long as forces are normal, too.

From Eq.~(\ref{eq:elaEnerg2D}) the elastic force onto the surface
layer can be derived.
In static equilibrium, it must be balanced by some external pressure.
This leads to the following equilibrium condition:
\begin{equation}
\frac{E^*}{2} q \tilde{u}({\bf q}) +
\tilde{p}_{\rm if}({\bf q}) + \tilde{p}_{\rm ext}({\bf q}) = 0,
\label{eq:stress}
\end{equation}
where
$\tilde{p}_{\rm if}({\bf q})$ is the interfacial force, e.g., the
Fourier transform of the pressure that the top solid exerts on the bottom
solid, and $\tilde{p}_{\rm ext}({\bf q})$ is an externally exerted
pressure.
For a constant external pressure, i.e.,
$\tilde{p}_{\rm ext}({\bf q}) = p_0 \,\delta_{0,{\bf q}}$
the equilibrium condition can be written as:
\begin{equation}
\tilde{p}_{\rm if}({\bf q}) =
\begin{cases}
-p_0 & \mbox{ if }{\bf q} = 0, \\
-\frac{E^*}{2} q \tilde{u}({\bf q}) & \mbox{else.}
\end{cases}
\end{equation}

Let $\tilde{u}_{\rm old}({\bf q})$ be a solution for a
given height profile $h({\bf r})$.
One can then construct a new solution for a system in which all
in-plane coordinates are scaled according to
$(x,y)_{\rm new} = s\cdot (x,y)_{\rm old}$,
which implies ${\bf q}_{\rm new} = {\bf q}_{\rm old}/s$:
\begin{equation}
\frac{E^*}{2}(s\cdot q_{\rm new})
\tilde{u}_{\rm old}(s\cdot {\bf q}_{\rm new}) =
-\tilde{p}_{\rm if, old}(s\cdot {\bf q}_{\rm new}).
\end{equation}
This equation can be reexpressed as
\begin{equation}
\frac{E^*}{2} q_{\rm new}
\tilde{u}_{\rm new}({\bf q}_{\rm new}) =
-\tilde{p}_{\rm if, new}( {\bf q}_{\rm new})
\end{equation}
with
\begin{equation}
\tilde{p}_{\rm if, new}({\bf q}_{\rm new}) =
\frac{1}{s}\cdot
\tilde{p}_{\rm if, old}({\bf q}_{\rm new}/s).
\end{equation}
Thus, all interfacial forces scale with $1/s$, which then must
also hold for the external pressure.

Our scaling transformation leaves $\bar{h}$ invariant and only
changes $\bar{g}$ to $\bar{g}/s$.
However, by renormalizing $p$ to $p/s$, we get back our
old solution.
Therefore, $p/E^*\bar{g}$ is the only variable which the mean separation
can depend on.
This simplifies Eq.~(\ref{eq:meanGapStructure}) to
\begin{equation}
\Delta \bar{u} = \bar{h} \cdot \Delta u_{\rm dl}(p/E^*\bar{g},H).
\end{equation}
For other observables, similar relationships can be found,
where the prefactor is a product of a power of $\bar{h}$
and the elastic constant $E^*$, and the relevant dimensionless
parameters are $p/E^*\bar{g}$ and $H$.

In some cases, one might find that $\Delta u_{\rm dl} = 0$.
This, however, does not imply that $\Delta \bar{u}$ is zero
when expressed as a multiple of a {\it microscopic} length.
It can still be finite, but since $\bar{h}$ can have diverged
in the fractal limit, the ratio $\Delta \bar{u}/\bar{h}$ has become zero.
In that case, it can be more appropriate to nondimensionalize
the gap --- or other quantities of unit length --- by a microscopic or mesoscopic
length.
One possibility to achieve this is the use of the following alternative
dimensionless expression for the mean gap $\Delta u_{\rm adl}$ via
\begin{equation}
\Delta \bar{u} = (\bar{g} \lambda_{\rm s}) \cdot \Delta u_{\rm adl}(p/E^*\bar{g},H).
\end{equation}
This is shown in Fig.~\ref{fig:gapFractalCorrLsunit} below.
Using Eqs.~(\ref{eq:scalCinRec}), (\ref{eq:barH}), (\ref{eq:barG}),
the leading-order term of
$\lambda_{\rm s}(\epsilon_{\rm f})$ can be written as
\begin{equation}
\bar{g} {\lambda_{\rm s}} = 2\pi \bar{h}
\left\{
 \begin{array}{ll}
 1/\sqrt{-2\ln \epsilon_{\rm f}}
 &\mbox{ } H=0, \vspace*{1mm} \\
 \sqrt{{H}/({1-H})} \epsilon_{\rm f}^H
 & \mbox{ } 0<H<1, \vspace*{1mm} \\
 \epsilon_{\rm f}\sqrt{-2\ln \epsilon_{\rm f}} & \mbox{ } H=1,
 \end{array}
\right.
\end{equation}
with $\epsilon_{\rm f} = \lambda_{\rm s}/\lambda_{\rm l} < 1$.

\section{Results}
\label{sec:results}
\subsection{Continuum, fractal, and finite size corrections}

\subsubsection{Contact area}

Any contact mechanics simulation is conducted at a finite system size ${\cal L}$
rather than in the thermodynamic limit.
The ratio $\lambda_{\rm l}/\lambda_{\rm s}$ is also finite
and the roughness at the smallest wavelength is discretized
only down to a finite ratio $a/\lambda_{\rm s}$.
In this section we investigate to what extent one can
express a dimensionless observable ${\cal O}_{\rm d}^{\mathrm{sim}}$
computed at finite values of
$\epsilon_{\rm t} = \lambda_{\rm l}/{\cal L}$,
$\epsilon_{\rm f} = \lambda_{\rm s}/\lambda_{\rm l}$, and
$\epsilon_{\rm c} = a/\lambda_{\rm s}$
through
\begin{align}
&{\cal O}_{\rm d}^{\mathrm{sim}}(p/E^*\bar{g},H,\epsilon_{\rm t},\epsilon_{\rm f},\epsilon_{\rm c}) = \notag \\
&{\cal O}_{\rm d}^{\mathrm{TFC}}(p/E^*\bar{g},H) +
C_{\rm t}  \epsilon_{\rm t}^{\alpha_{\rm t}} +
C_{\rm f}  \epsilon_{\rm f}^{\alpha_{\rm f}} +
C_{\rm c}  \epsilon_{\rm c}^{\alpha_{\rm c}}
\label{eq:tfc}
\end{align}
and extrapolate to the TFC limit by computing the observable of
interest at finite values of $\epsilon_{\rm t}$, $\epsilon_{\rm f}$,
and $\epsilon_{\rm c}$.
Both the exponents $\alpha_i$ and also the proportionality constants
$C_i$ can be functions of $H$ and $p/E^*\bar{g}$, though one would expect
the exponents to depend only weakly on the pressure.

Note that any of the three corrections \textit{can be significant}
and it is not \textit{a priori} clear which correction is important for a given
combination of $H$ and $p/E^*\bar{g}$.
Therefore, it is important to systematically control the values of the parameters
independently of each other, i.e., by only changing one $\epsilon_i$ at
a time.
This point is taken into account in our calculations in the following way:
we choose a reference system, namely,
${\cal L}/\lambda_{\rm l} = 2$, $\lambda_{\rm l}/\lambda_{\rm s} = 1024$,
$\lambda_{\rm s}/a = 2$, and run simulations by varying each value
of $\epsilon_i$ while keeping the other two constant.
To estimate stochastic error bars, we perform calculations with up to four
different realizations of the randomly rough surfaces.

All the calculations have been carried out using the GFMD technique described in Appendix~\ref{sec:GFMD} and implemented in our in-house parallel code. Note that to get reliable extrapolated results in one of the limits, it is necessary to approach the limit as closely as possible. This requires considering quite large system sizes ${\cal L}$. The typical ``close to a limit'' system size in the present study is ${\cal L} = 2^{15}$ which corresponds to about 1 billion grid points in the 2D contact plane. The largest linear size presented here is ${\cal L} = 2^{17}$ (about 17 billion grid points in the contact plane) which we used to ascertain the scaling of the mean gap with $\epsilon_{\rm f}$ (see Fig.~\ref{fig:gapFractalCorr} below).

We start our analysis with the continuum corrections to the contact area,
which historically were the first ones to come under scrutiny.
Hyun and Robbins~\cite{Hyun04} evaluated contact area using $\epsilon_{\rm c} = 1/2$
arguing that roughness extends down to the smallest scale.
This argument is valid if one is interested in determining the
contact area of real systems, although a rigorous definition of
contact may only be possible in the realm of continuum mechanics.
We therefore feel that the limit $\epsilon_{\rm c} \to 0$ is more
appropriate for our purpose,
all the more when testing the validity of a solution in
continuum mechanics. To this date, no consensus has been reached to the precise value of the
dimensionless ratio $\kappa \equiv a_{\mathrm{r}}\bar{g}E^*/p$  for small $p$.
Recent numerical estimates range from $\kappa = 2$ ($\epsilon_{\rm c} = 1/32$,
$\epsilon_{\rm f} = 1/64$, $\epsilon_{\rm t} = 1$) in Ref.~\cite{Putignano12} to
values exceeding 2.5
(e.g., $\epsilon_{\rm c} = 1/16$, $\epsilon_{\rm f} = 1/8$,
$\epsilon_{\rm t} = 1/16$) in Ref.~\cite{Yastrebov12},
while Persson theory predicts that $\kappa = \sqrt{8/\pi}\approx 1.6$.

\begin{figure}[htb]
\includegraphics[width=0.45\textwidth]{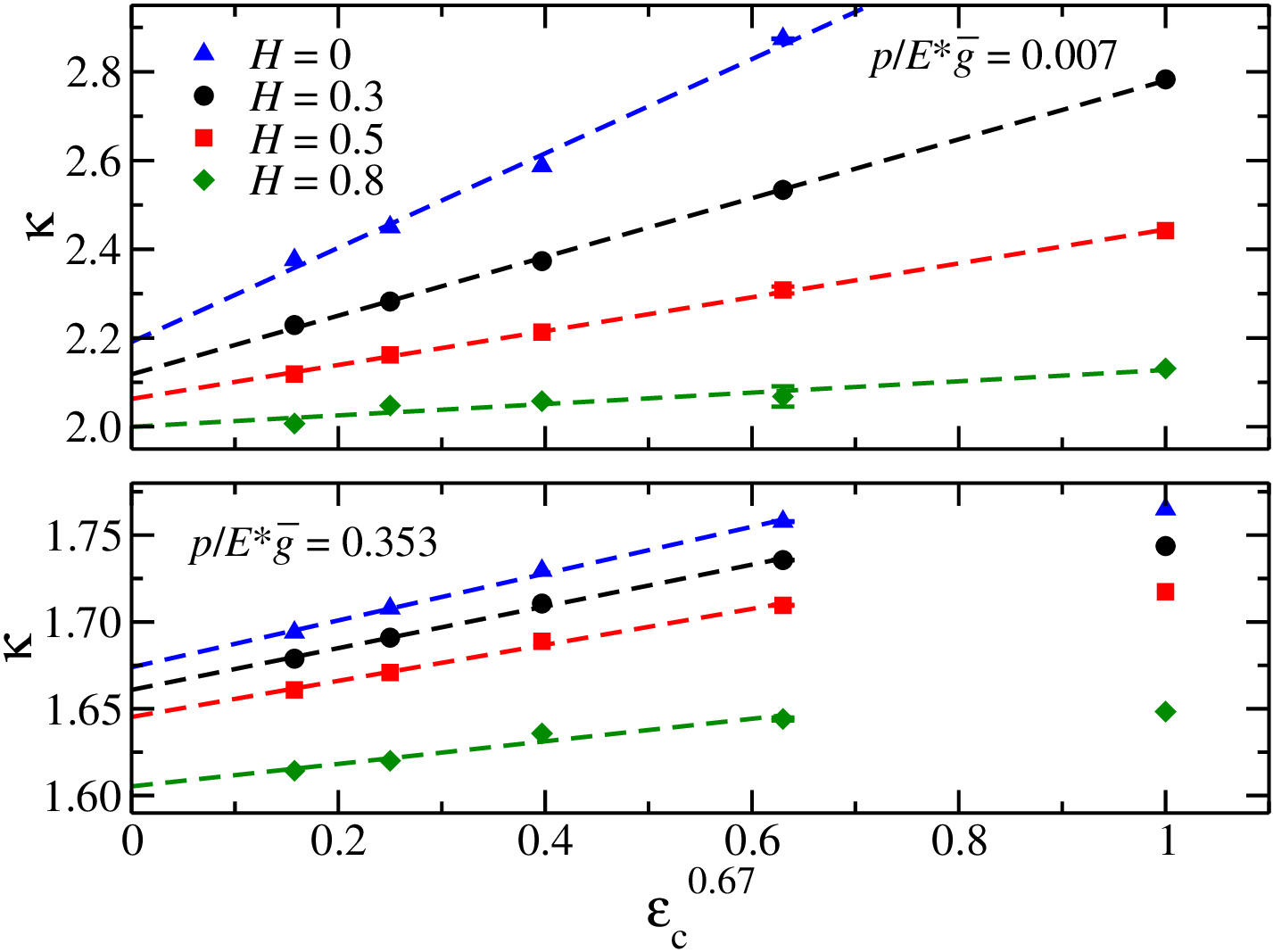}
\caption{
Proportionality coefficient $\kappa$
as a function of the discretization $\epsilon_{\rm c}$
for fixed values $\epsilon_{\rm t} = 1/2$ and $\epsilon_{\rm f} = 1/1024$.
One set of curves is evaluated at a low pressure $p^* = p/E^*\bar{g} = 0.007$
(top panel),
while another one is evaluated at $p^* = 0.353$
(bottom panel).
\label{fig:continuumCorr}
}
\end{figure}

In Fig.~\ref{fig:continuumCorr} we show how the proportionality
coefficient $\kappa$ depends on $\epsilon_{\rm c}$ for fixed values of $\epsilon_{\rm f} = 1/1024$
and $\epsilon_{\rm t} = 1/2$.
Here, $\kappa$ is defined as
\begin{equation}
\kappa = \frac{A}{A_0 p^*},
\end{equation}
where $A$ and $A_0$ are  real and apparent contact area, respectively,
and $p^* = p/E^*\bar{g}$.
We find that the $\epsilon_{\rm c}$ correction follows a power law with the
exponent $\alpha_{\rm c} \approx 0.67$ at small loads.
This is in agreement with the work by Campa\~{n}\'{a} and M\"user~\cite{Campana07} who found the same exponent, although their work was not yet based
on the continuum Green's functions but rather on
the Green's functions describing a {\it discrete} elastic manifold.
Other differences are that Campa\~{n}\'{a} and M\"user~\cite{Campana07}
did not keep $\epsilon_{\rm f}$ constant and used
$\epsilon_{\rm t} = 1$.
Despite these distinctions, we confirm that $\kappa$ in the continuum limit is indeed close to 2
and  that it increases marginally as $H$ decreases.
This is also consistent  with Putignano \textit{et al.} who found virtually no dependence
on $H$ for values of $H$ close to unity~\cite{Putignano12,Putignano13}, and also
with Persson theory in which $\kappa$ is independent of $H$.

While previous work focused on the low-load limit, we extend
the analysis of discretization corrections to larger pressures,
where the $A \propto p$ is no longer accurate.
In this regime $\kappa$ falls below the value of 2.
The contact area over pressure still appears to converge  with $\epsilon^{0.67}_{\rm c}$.
The corrections are smaller, and convergence starts at smaller values
of $\epsilon_{\rm f}$ than at lower pressures.
It is interesting to note that the relative contact area still seems to be
rather independent of $H$.
At the given values of $\epsilon_{\rm f}$ and $\epsilon_{\rm t}$,
its value ranges from 0.837 for $H = 0.8$ to 0.802 for $H = 0$ at $p^* = 0.353$.

\begin{figure}[htb]
\includegraphics[width=0.45\textwidth]{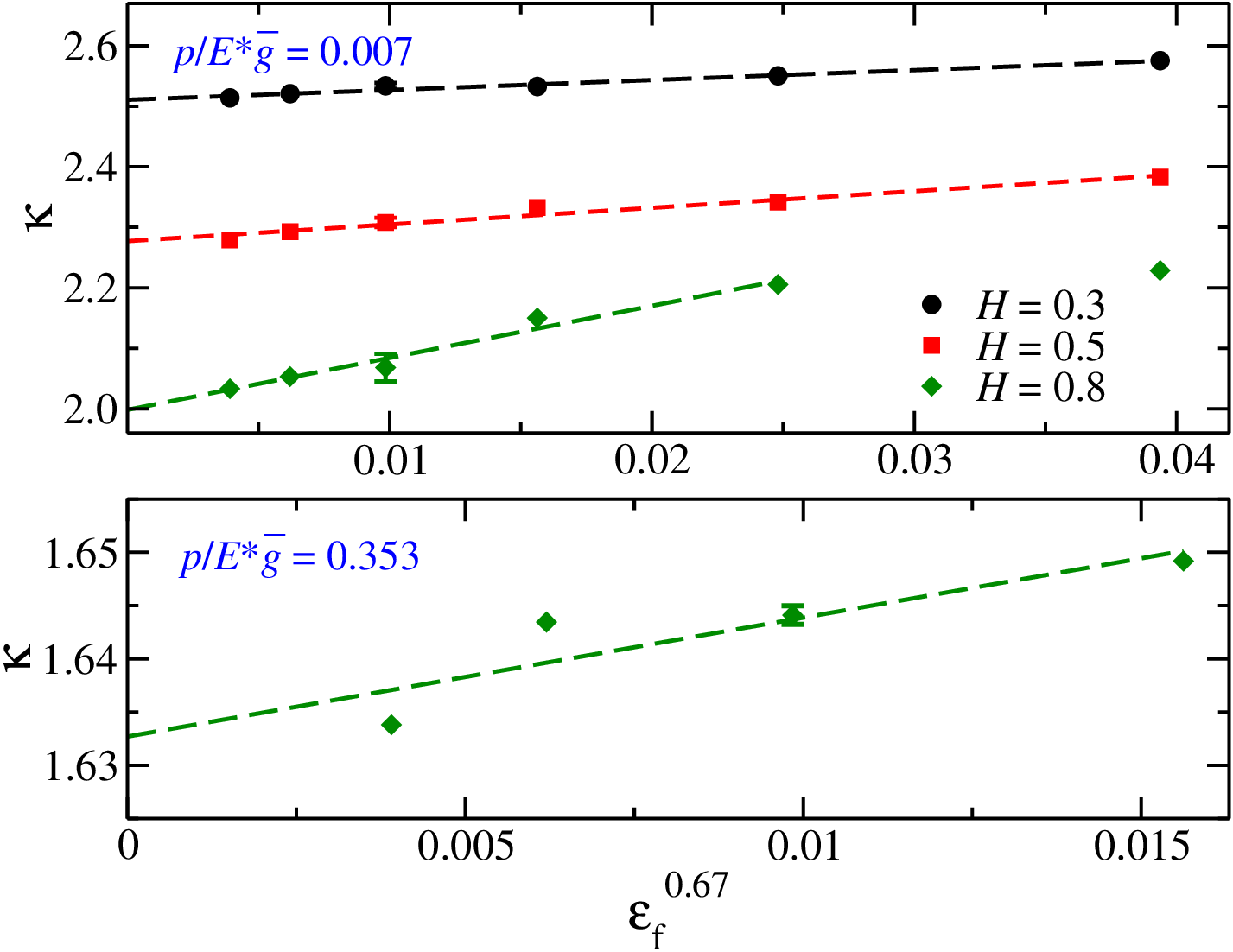}
\caption{
Proportionality coefficient $\kappa$
as a function of the fractal correction
$\epsilon_{\rm f} = \lambda_{\rm s}/\lambda_{\rm l}$
for fixed values $\epsilon_{\rm c} = 1/2$ and $\epsilon_{\rm t} = 1/2$.
One set of curves is obtained at a low pressure $p^* = p/E^*\bar{g} = 0.007$,
while another one is evaluated at $p^* = 0.353$. At higher pressure only one set of data ($H=0.8$) is shown to emphasize a small value of the fractal correction. For other values of $H$ the effect of $\epsilon_{\rm f}$ is even smaller.
\label{fig:fractalCorr}
}
\end{figure}

Figure~\ref{fig:fractalCorr} depicts the influence of the fractal
correction $\epsilon_{\rm f}$ on $\kappa$ for two pressures
at the default reference values of $\epsilon_{\rm c}$ and $\epsilon_{\rm t}$.
For both pressures, we find an exponent of $\alpha_{\rm f} \approx 0.67$.
One can see that larger values of $\epsilon_{\rm f}$ can lead to
substantial errors in particular at small pressures and Hurst exponents close to 1.
For example, on the $p^* = 0.007$ curve for $H = 0.8$ the error
initially increases with $ \approx 3 \epsilon_{\rm f}^{0.67}$ so that
to a zeroth-order approximation, $\kappa$ may be overestimated by
as much as ${\cal O}(30\%)$ if $\lambda_{\rm l}/\lambda_{\rm s} = 32$ is
chosen.
This might explain why recent work by Yastrebov{\it~et al.}~\cite{Yastrebov12},
who focused on continuum corrections, found particularly large values
for $\kappa$ in contrast to studies~\cite{Campana07,Putignano12}
employing smaller values of $\epsilon_{\rm f}$.
The problem appears to be that if one keeps ${\lambda_{\rm l}}/a$ constant
but varies $\lambda_{\rm s}$, it might remain unnoticed that the error is
converted from a continuum correction to a fractal correction, all
the more as both corrections are positive and each scales only
sublinearly with $\epsilon$.

It may also be interesting to note that
Yastrebov{\it et al.}~\cite{Yastrebov12} found values for $\kappa$ close to
the predictions based on the Bush-Gibson-Thomas (BGT) theory~\cite{Persson06}, which
is an asperity-based model neglecting that individual contact
patches can merge to form fractal shaped contact patches.
This suggests that systems with small values of $\epsilon_{\rm f}$
do not always behave like self-affine randomly rough surfaces but more
like the collection of (elastically uncoupled) bumps.

For larger values of $p^*$ we again find small prefactors for the
corrections to $\kappa$.
Within the stochastic scatter, leading-order corrections are
consistent with an ${\cal O}(\epsilon_{\rm f}^{0.67})$ dependence.
However, prefactors are small.
In the case of $H = 0.3$ the fractal corrections are  even
below our (statistical) detection capabilities.
Despite the small prefactors, it seems as if
convergence starts at smaller values of
$\epsilon_{\rm f}$ than for smaller pressures.

\begin{figure}[htb]
\includegraphics[width=0.45\textwidth]{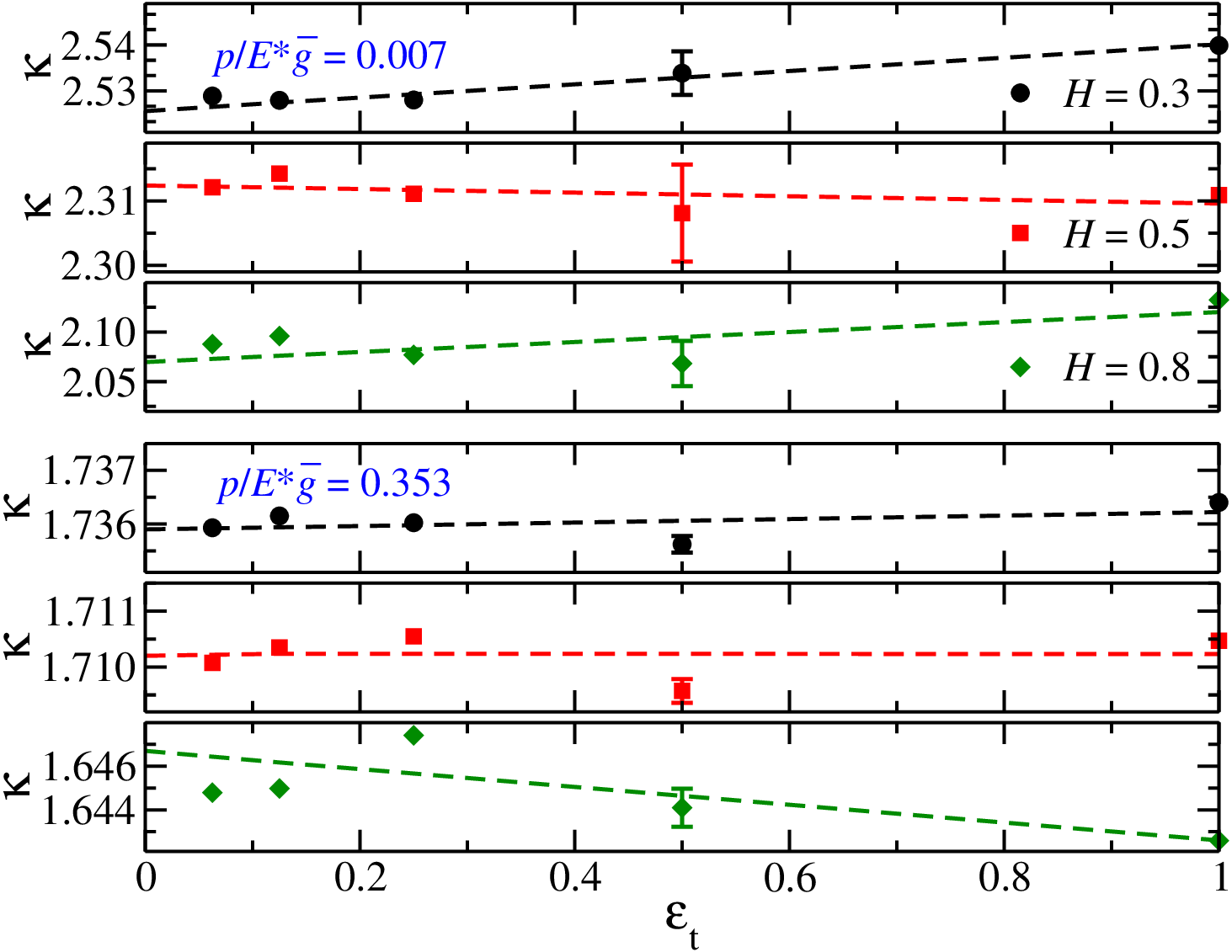}
\caption{
Dimensionless ratio $\kappa$
as a function of the thermodynamic correction $\epsilon_{\rm t}$
for fixed values $\epsilon_{\rm c} = 1/2$ and $\epsilon_{\rm f} = 1/1024$.
Dotted lines indicate linear fits. $\kappa$ is nearly independent of the thermodynamic correction.
\label{fig:thermodynCorr}
}
\end{figure}

In the study of TFC corrections to the contact area, we also investigate
how the thermodynamic limit is approached.
Figure~\ref{fig:thermodynCorr} reveals that the dependence of $\kappa$
on $\epsilon_{\rm t}$ is rather weak.
In fact, corrections are so small that we cannot determine with
certainty the exponent $\alpha_{\rm t}$.
The fluctuations in $\kappa$ are of order 1\% and 0.1\% for $p^* = 0.007$ and $p^* = 0.353$, respectively.

With regards to finding the TFC limit for the contact area,
we summarize that continuum and fractal
corrections are quite substantial, while those related to the
thermodynamic limit are minor.
The FC corrections have a larger prefactor at smaller
pressures; however, the asymptotic convergence to the limits starts at larger values of $\epsilon_{\rm c}$ and
$\epsilon_{\rm f}$ than at higher pressures.
As expected, fine discretizations (small $\epsilon_{\rm c}$)
are required for small values of $H$, where roughness lives more
strongly on short wavelengths than at large values of $H$.
Conversely, one must ensure relatively small values of $\epsilon_{\rm f}$
for large $H$.

From the GFMD results we obtained Tables~\ref{table:kappaLowP} and \ref{table:kappaHighP} containing coefficients and powers to be used in Eq.~(\ref{eq:tfc}). This information allows extrapolating $\kappa$ to the TFC limit. As a specific example, to extrapolate $\kappa^{\mathrm{sim}}$ at $p^* = 0.007$ and $H = 0.8$, one can add the computed
corrections according to Eq.~(\ref{eq:tfc}) for the values of $\epsilon_{\rm t}$,
$\epsilon_{\rm f}$ and $\epsilon_{\rm c}$ used in the simulations:
\begin{align}
&\kappa^{\mathrm{TFC}}(p^* = 0.007,H = 0.8) \approx \kappa^{\mathrm{sim}}(p^*,H, \epsilon_{\rm t}, \epsilon_{\rm f}, \epsilon_{\rm c}) \notag \\
&-0.0421  \epsilon_{\rm t} -
8.5879  \epsilon_{\rm f}^{0.67} -
0.1279  \epsilon_{\rm c}^{0.67}.
\label{eq:tfcArea}
\end{align}

\begin{table}[htb]
\begin{tabular}{c|c|c|c|c|c|c}
  $H$ & $C_{\mathrm{c}}$ &$C_{\mathrm{f}}$ & $C_{\mathrm{t}}$ & $\alpha_{\mathrm{c}}$ & $\alpha_{\mathrm{f}}$ & $\alpha_{\mathrm{t}}$ \\ \hline
  0.3 & 0.6621 & 1.6322 & 0.0132 & 0.67 & 0.67 & 1 \\
  0.5 & 0.3817 & 2.7518 & -0.0028 & 0.67 & 0.67 & 1 \\
  0.8 & 0.1279 & 8.5879 & 0.0421 & 0.67 & 0.67 & 1 \\
\end{tabular}
\caption{Coefficients and powers in Eq.~(\ref{eq:tfc}) for $\kappa$ at $p^* = 0.007$ obtained from the GFMD results.}
\label{table:kappaLowP}
\end{table}

\begin{table}[htb]
\begin{tabular}{c|c|c|c|c|c|c}
  $H$ & $C_{\mathrm{c}}$ &$C_{\mathrm{f}}$ & $C_{\mathrm{t}}$ & $\alpha_{\mathrm{c}}$ & $\alpha_{\mathrm{f}}$ & $\alpha_{\mathrm{t}}$ \\ \hline
  0.3 & 0.1202 & 0.0535 & $3.1658\cdot10^{-4}$ & 0.67 & 0.67 & 1 \\
  0.5 & 0.1037 & 0.0796 & $8.4447\cdot10^{-5}$ & 0.67 & 0.67 & 1 \\
  0.8 & 0.0650 & 1.1145 & $-3.1623\cdot10^{-3}$ & 0.67 & 0.67 & 1 \\  
\end{tabular}
\caption{Coefficients and powers in Eq.~(\ref{eq:tfc}) for $\kappa$ at $p^* = 0.353$ obtained from the GFMD results.}
\label{table:kappaHighP}
\end{table}

Unfortunately, it is difficult to predict for what range of pressures the data for
$p^* = 0.007$ can be used without considerable loss of accuracy.
Although $a_{\mathrm{r}}$ is linear in $p^*$ at low $p^*$ ($\lesssim 10^{-1}$),
one averages over a different distribution of contact patches when decreasing $p^*$,
e.g., the largest contact patch shrinks with decreasing $p^*$, and the relative
importance of small contact patches increases.
Thus, our tables only convey trends as to which corrections become important
for different $H$ in the high- or low-pressure regime. 

Despite potentially large corrections, well-chosen parameters
allow one to produce quite meaningful results, even without extrapolation.
For example, for
$H = 0.8$ and $p^* = 0.007$ if one chooses $a/\lambda_{\rm s} = 1/2$, $\lambda_{\rm s}/\lambda_{\rm l} = 1/1024$, and $\lambda_{\rm l}/{\cal L} = 1$, the expected error in $\kappa$ is only $6\%$ (for a system size of 2048).
However, to keep the error for a system with $H = 0.3$ similarly small, it is better to use  $a/\lambda_{\rm s} = 1/16$ and $\lambda_{\rm s}/\lambda_{\rm l} = 1/256$.
A system of total size $2048\times 2048$ can be easily handled on a single CPU core,
and convergence with a well-tuned GFMD code is reached within less than
an hour of computing time on modern hardware.

\subsubsection{Mean gap}

Unlike the contact area, the TFC analysis of the mean gap $\Delta \bar{u}$ has not attracted much attention in the literature.
Therefore, there are not many references for the $\Delta \bar{u}$ to compare with (in contrast to $a_{\mathrm{r}}$ or $\kappa$ for which there exist computational results from several research groups). Additionally, the TFC analysis for the mean gap is more challenging
than that for $a_{\mathrm{r}}$. As was mentioned in Section~\ref{sec:dimAnal} and as will be clear from the GFMD results, the choice of the unit for the length scale considerably influences the value of $\Delta \bar{u}$ in the fractal limit. However, when the continuum and thermodynamic corrections are considered, the question about the unit of length is not very important and we use the RMS height $\bar{h}$ for normalization in these cases. Note that we also compare the GFMD results with Persson theory, which can predict scaling of the $\Delta \bar{u}$ with the fractal correction (see the Appendix~\ref{append:persson} for more information on Persson theory). For other two corrections such a comparison is not possible as the theory assumes that the continuum and thermodynamic limits are satisfied.

\begin{figure}[htb]
\includegraphics[width=0.45\textwidth]{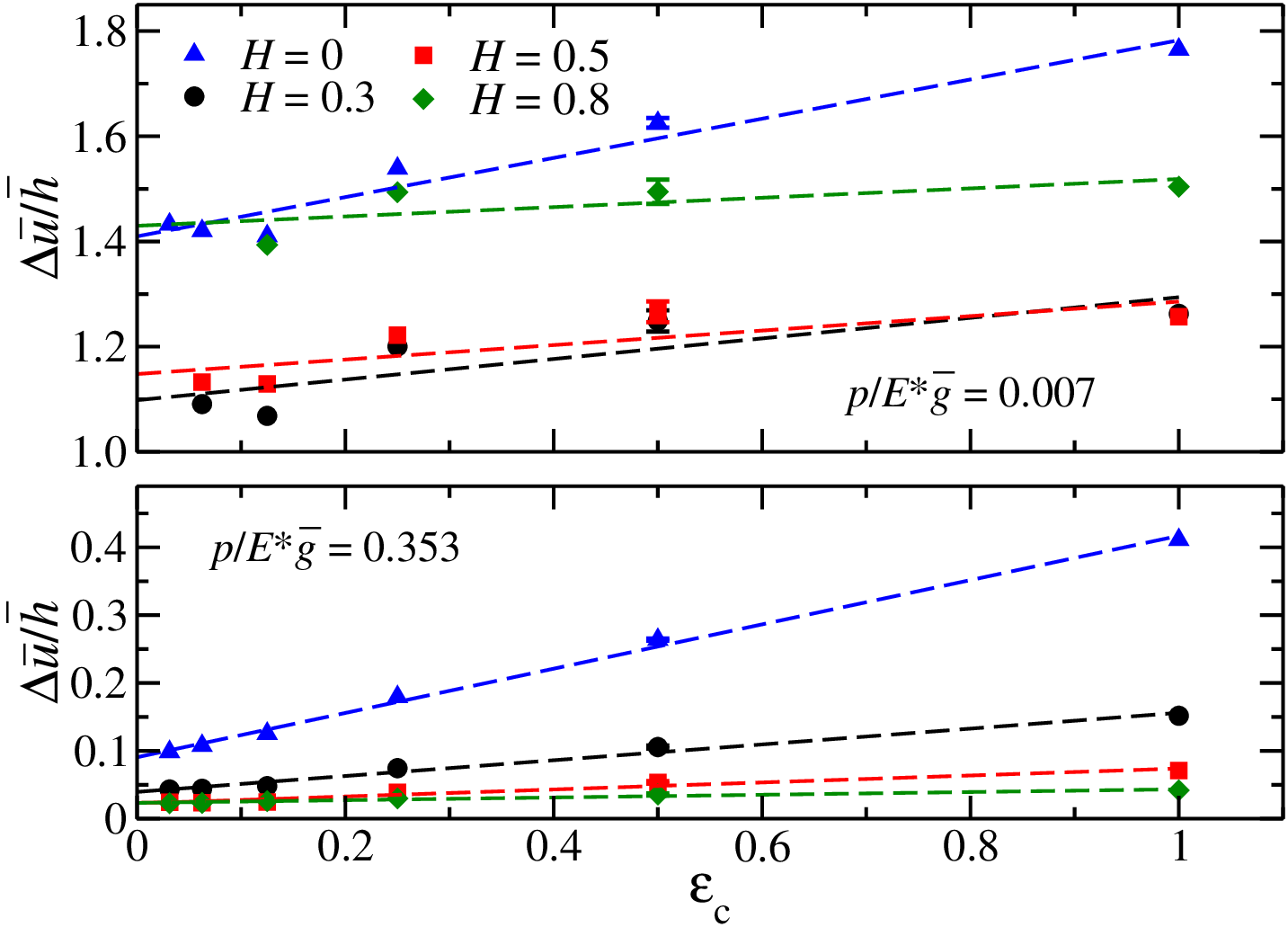}
\caption{
Dimensionless gap in units of the RMS height as a function of the discretization $\epsilon_{\rm c}$ for
fixed $\epsilon_{\rm t} = 1/2$ and $\epsilon_{\rm f} = 1/1024$.
One set of curves is evaluated at a low pressure $p^* = p/E^*\bar{g} = 0.007$
(top panel), while another one is evaluated at $p^* = 0.353$
(bottom panel). Dashed lines indicate least-squares linear fits.
\label{fig:gapContinuumCorr}
}
\end{figure}

As before, the continuum limit is examined first. From Fig.~\ref{fig:gapContinuumCorr} one can see that the discretization corrections are largest for the smallest values of $H$. At high pressures the corrections follow a linear law. At the lower pressure, numerical noise does not allow one to establish a scaling law. Nevertheless, the linear fits can again be used with high accuracy. Prefactors can be large; for $p^* = 0.353$ and $H=0$, the gap is almost three times larger for $\epsilon_{\rm c} = 1/2$ (which is the reference value) than in the (extrapolated) limit of $\epsilon_{\rm c} = 0$.
In contrast, the corrections for $H= 0.8$ are relatively moderate, e.g., they only amount to
roughly 20\% of the observable at $\epsilon_{\rm c} = 1/2$.
For this larger value of $H$, we are plagued with large stochastic error at small $p$,
which is readily seen from the top panel of Fig.~\ref{fig:gapContinuumCorr}.

\begin{figure}[htb]
\includegraphics[width=0.45\textwidth]{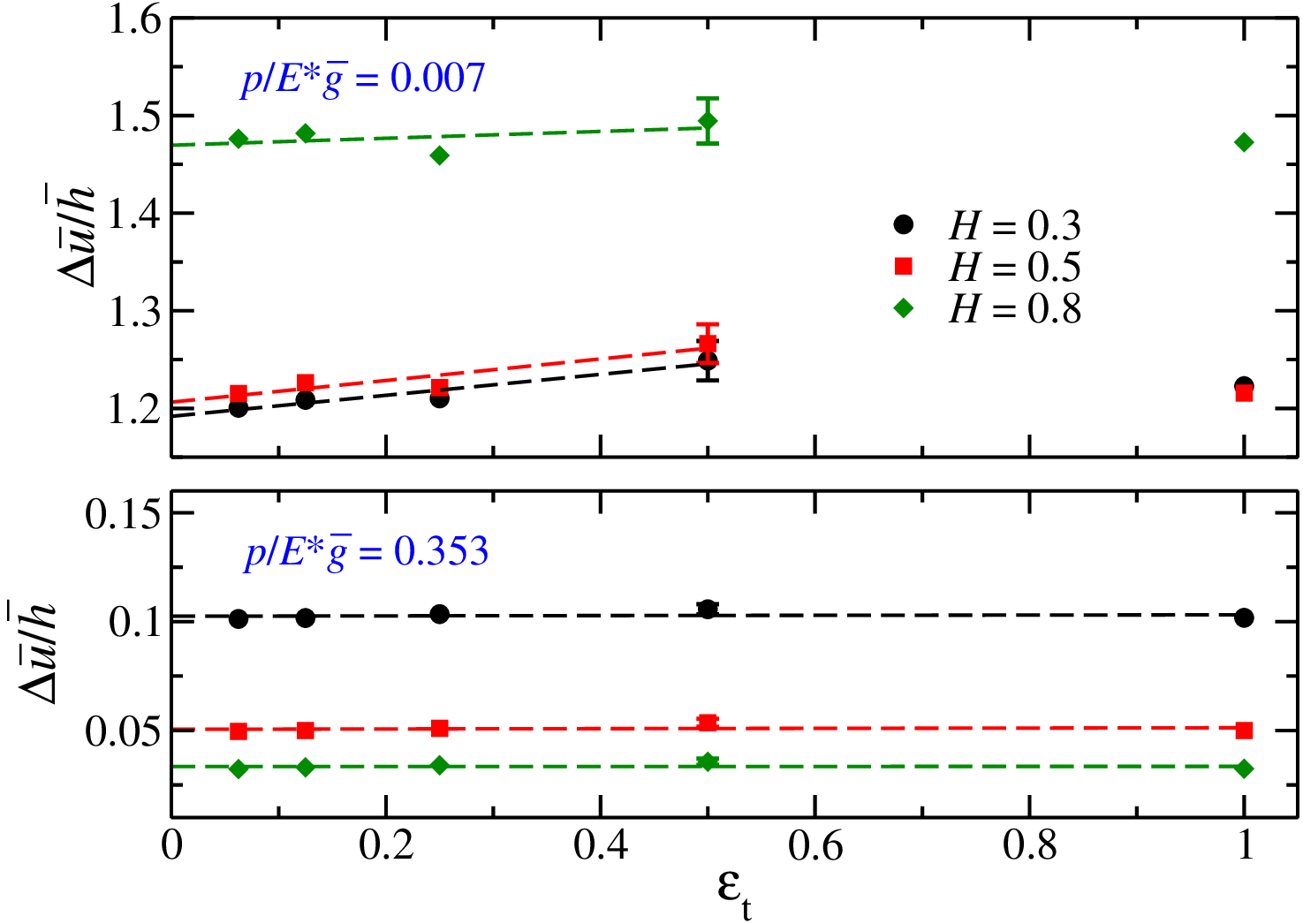}
\caption{
Dimensionless gap in units of the RMS height as a function of the thermodynamic correction $\epsilon_{\rm t}$
for fixed values $\epsilon_{\rm c} = 1/2$ and $\epsilon_{\rm f} = 1/1024$.
Dashed lines indicate least-squares linear fits.
\label{fig:gapThermodynCorr}
}
\end{figure}

Thermodynamic corrections show similar trends as continuum corrections, as one
can see in Fig.~\ref{fig:gapThermodynCorr}.
Specifically, we find that $\Delta \bar{u}$ varies linearly in $\epsilon_{\rm t}$. The numerical scatter is again particularly large for low $p$ and larger $H$.

\begin{figure}[htb]
\includegraphics[width=0.45\textwidth]{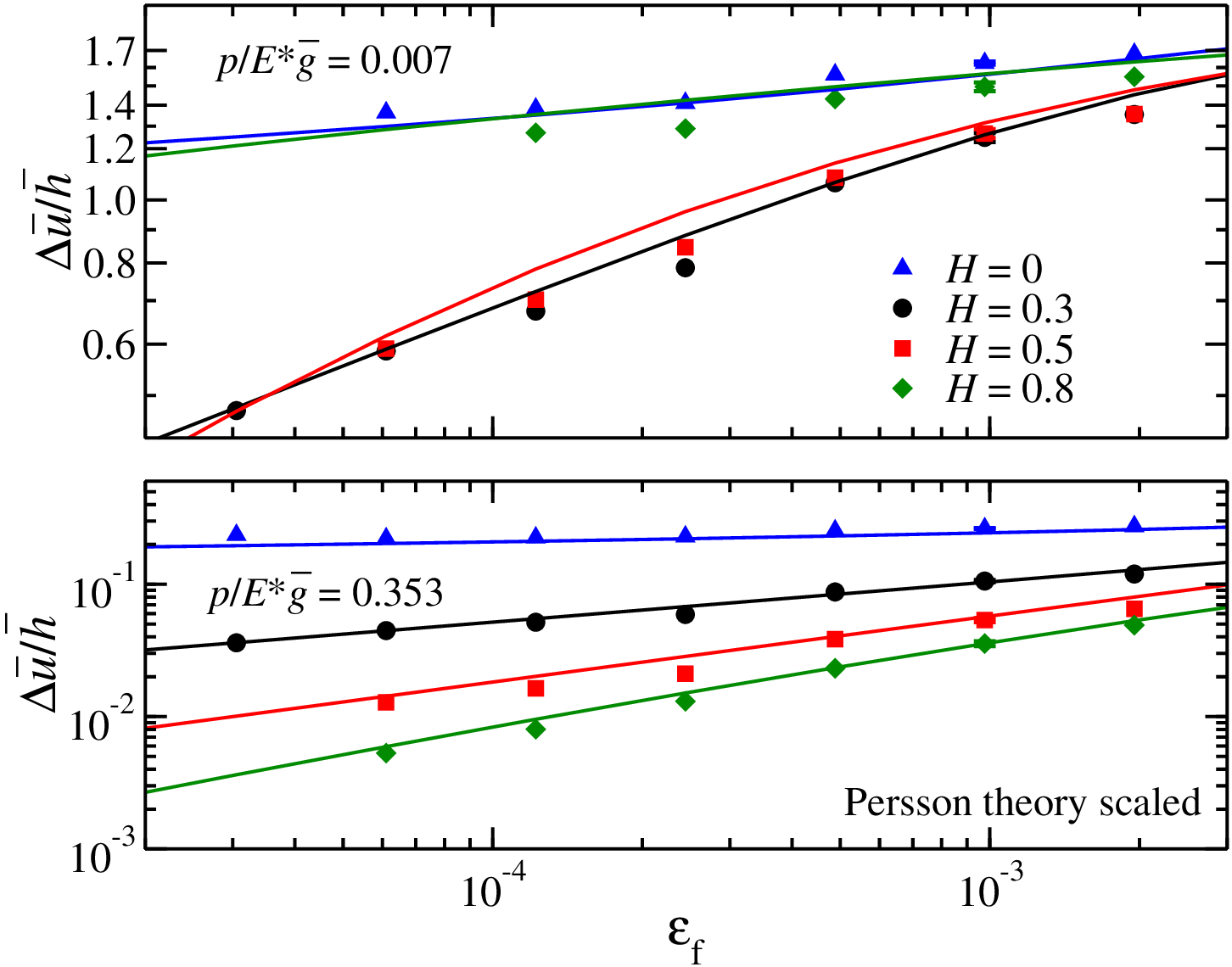}
\caption{
Dimensionless gap in units of the RMS height as a function of the fractal
correction $\epsilon_{\rm f} = \lambda_{\rm s}/\lambda_{\rm l}$
for fixed values $\epsilon_{\rm c} = 1/2$ and $\epsilon_{\rm t} = 1/2$.
One set of curves is evaluated at a low pressure $p^* = p/E^*\bar{g} = 0.007$
(top panel), while another one is evaluated at $p^* = 0.353$
(bottom panel). Dots show GFMD results, lines are obtained using Persson theory. To match with the simulations within 10\% at the higher pressure theory results are scaled by factors of 3 and 2 for $H=\{0, 0.3, 0.5\}$ and $H=0.8$, respectively. No scaling is required at the lower pressure.
\label{fig:gapFractalCorr}
}
\end{figure}

\begin{figure}[htb]
\includegraphics[width=0.45\textwidth]{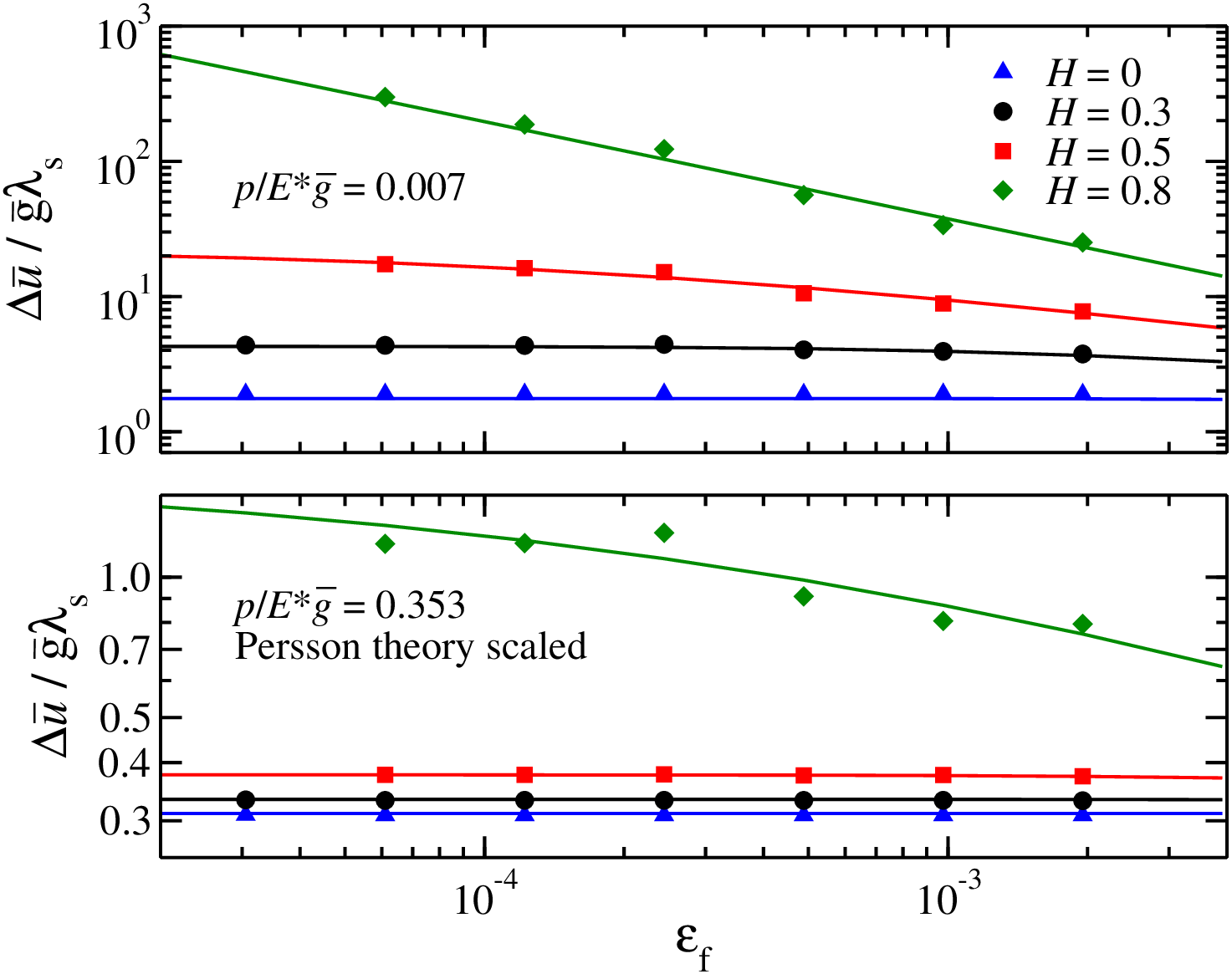}
\caption{
Dimensionless gap in units of $\bar{g}\lambda_{\rm s}$ as a function of the fractal
correction $\epsilon_{\rm f} = \lambda_{\rm s}/\lambda_{\rm l}$
for fixed values $\epsilon_{\rm c} = 1/2$ and $\epsilon_{\rm t} = 1/2$. The focus is on the larger values of $\epsilon_{\rm f}$ captured by the GFMD.
Dots show GFMD results, lines are obtained using Persson theory. To match with the simulations within 10\% at the higher pressure theory results are scaled by factors of \{3.3, 3.1, 2.75, 2\} for $H=\{0, 0.3, 0.5,0.8\}$, respectively. No scaling is required at the lower pressure.
\label{fig:gapFractalCorrLsunit}
}
\end{figure}

\begin{figure}[htb]
\includegraphics[width=0.45\textwidth]{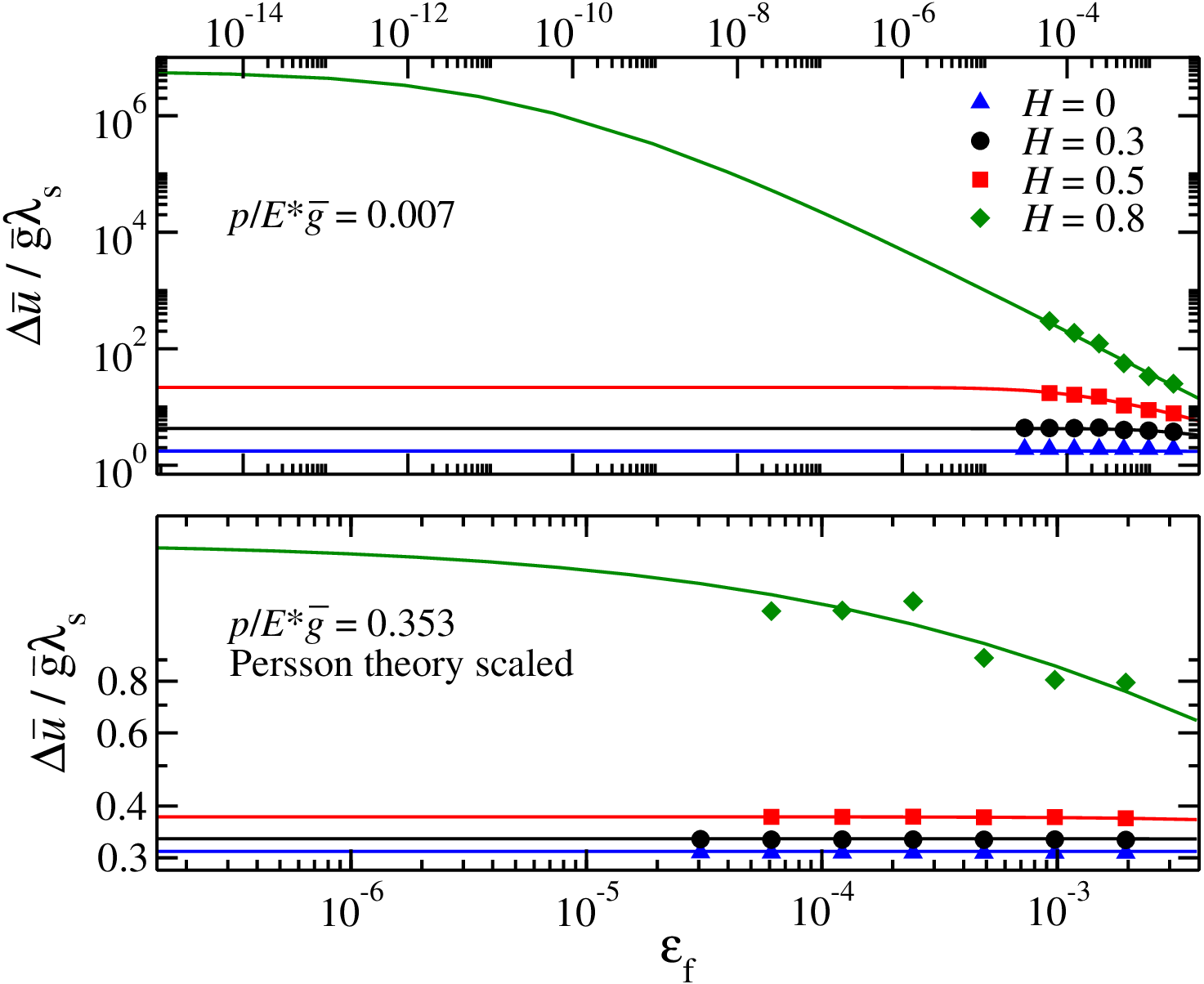}
\caption{
Dimensionless gap in units of $\bar{g}\lambda_{\rm s}$ as a function of the fractal
correction $\epsilon_{\rm f} = \lambda_{\rm s}/\lambda_{\rm l}$
for fixed values $\epsilon_{\rm c} = 1/2$ and $\epsilon_{\rm t} = 1/2$. Values of $\epsilon_{\rm f}$ span almost 15 decades. According to the theory, the mean gap eventually converges in the fractal limit.
Designations and scaling are the same as in Fig.~\ref{fig:gapFractalCorrLsunit}.
\label{fig:gapFractalCorrLsunitConverged}
}
\end{figure}

Fractal corrections obtained with GFMD and Persson theory are shown in Figs.~\ref{fig:gapFractalCorr}, \ref{fig:gapFractalCorrLsunit} and \ref{fig:gapFractalCorrLsunitConverged}. At the lower pressure, theory and simulations fit within about 10\% accuracy. Although the theory gives correct functional dependencies at the higher pressure, the prefactors are about 2--3 times smaller than in the simulations (see captions of Figs.~\ref{fig:gapFractalCorr}, \ref{fig:gapFractalCorrLsunit} for the exact values of the scaling factors).
Independent of the normalization, there exists a region where the mean gap varies with $\epsilon_{\rm f}$ approximately according to a power
law with pressure-dependent exponents. When normalizing with the RMS height $\bar{h}$, the exponents
at $p^*=0.007$ are $\alpha_{\rm f} = \{0.067,$ $0.26, 0.25, 0.079\}$ for $H = \{0, 0.3, 0.5, 0.8\}$ respectively. At $p^*=0.353$
the exponents are $\alpha_{\rm f} = \{0.048, 0.30, 0.51,$ $0.67\}$ for $H = \{0, 0.3, 0.5, 0.8\}$ respectively.
Using $\bar{g}\lambda_{\rm s}$ as the normalization factor leads to the increase of the mean gap with
$\epsilon_{\rm f}$ according to power laws with almost zero exponents for $H = \{0, 0.3\}$ both at
higher and lower pressures. The exponents for the cases of $H = \{0.5, 0.8\}$ are
$\alpha_{\rm f} = \{0.25, 0.75\}$ at $p^*=0.007$ and $\alpha_{\rm f} = \{0.002, 0.14\}$ at $p^*=0.353$.
These results imply that the mean gap in the fractal limit is zero in units of $\bar{h}$ at relatively
small pressures, although the gap measured
in microscopic units such as $\lambda_{\rm s}$ must be greater than zero
as long as contact is not complete.

From Figs.~\ref{fig:gapFractalCorr}--\ref{fig:gapFractalCorrLsunitConverged} one can learn that
extrapolating to the fractal limit is far from trivial.
One reason is that when $\epsilon_{\rm f} \rightarrow 0$, $\Delta \bar{u}$ expressed in units of $\bar{h}$ does not show a
clear plateau, while for $\bar{g}\lambda_{\rm s}$ extremely small values of $\epsilon_{\rm f}$ are required in order to see the convergence. Having only smaller (though practically relevant) values of $\epsilon_{\rm f}$ may lead to the incorrect conclusion that $\Delta \bar{u}/(\bar{g}\lambda_{\rm s})$ diverges according to a power law. In fact, even though we are using systems sizes up to ${\cal L} = 2^{17}$ spanning 5 orders of magnitude in length, we are still in the ``large $\epsilon_{\rm f}$ regime'' where we only see the power law divergence (Fig.~\ref{fig:gapFractalCorrLsunit}). Nevertheless, the theory predictions --- that agree well with the simulations at larger $\epsilon_{\rm f}$ --- clearly indicate the eventual convergence of $\Delta \bar{u}/(\bar{g}\lambda_{\rm s})$ to the fractal limit (Fig.~\ref{fig:gapFractalCorrLsunitConverged}).
These observations may also suggest that for the mean gap, $\epsilon_{\rm f}$ should
be considered as an additional independent variable and not as a correction,
in contrast to the situation for relative contact area.

Let us discuss in more detail the implications
of finding either (i) a finite value of
$\Delta \bar{u}/\bar{h}$ or (ii) a finite value of $\Delta \bar{u}/\bar{g}\lambda_{\rm s}$.
Observation (i) goes hand in hand with having only one or very few mesoscale
asperity contacts per area of size $\lambda_{\rm l}^2$.
Those contacts are found in the vicinity of the
highest peak on a domain of size $\lambda_{\rm l}^2$.
If contact is distributed more or less homogeneously throughout the apparent contact area,
gaps must automatically disappear when expressed in units of $\bar{h}$.
In this case, i.e., for pressures so small that contact occurs only near
the highest asperity, the original Persson theory presented in this work
is inappropriate and finite-size corrections need to be applied~\cite{Pastewka13}.

Observation (ii) implies contacts that start to look
homogeneous when spatial features on length scales only slightly larger
than $\lambda_{\rm s}$ are resolved.
In fact, given Eq.~(\ref{eq:scalGradient}),
one could argue that the root-mean-square height
due to the roughness on the shortest wavelengths is of order
$\sqrt{2-2H}\bar{g}\lambda_{\rm s}$.
Then, the gaps for large pressures
($p^* = 0.353$ and $H = \{0, 0.3, 0.5\}$
in Fig.~\ref{fig:gapFractalCorrLsunit})
are of order of (but smaller than) the
root-mean-square height associated with  short wavelengths.
Thus, non-trivial scaling of the gap with $\epsilon_{\rm f}$ can
occur when the contacts start to look heterogeneous at wavelengths
much less than $\lambda_{\rm l}$ but distinctly more
than $\lambda_{\rm s}$.

We conclude this section by identifying reasonable reference systems
for the evaluation of the mean gap.
For the $H = 0.8$, $p^* = 0.007$ system, discretization corrections
appear to be rather minor.
Choosing $a/\lambda_{\rm s}$ as large as 1/2 does not seem to introduce artifacts.
Finite-size corrections also do not lead to considerable errors.
Choosing $\lambda_{\rm l}/{\cal L} = 1/4$ makes the estimated error
for the mean gap be less than 10\%.
Lastly, the ratio $\lambda_{\rm s}/\lambda_{\rm l}$ has to be sufficiently
small, i.e., below 1/1024.
Thus, for such a system, a calculation should be as large as
$8192\times8192$ to achieve an accuracy of ${\cal O}(10\%)$
(in the absence of extrapolation).
This is the largest system that is commonly run on a
single commodity CPU core or on a single GPU.
The choice for the reference system for $H = 0$, $p^* = 0.353$
differs quite substantially from that just discussed.
Now, one would probably be better off with
$a/\lambda_{\rm s} = 1/16$. However, this time
${\cal L}/\lambda_{\rm s} = 1/512$ is more than sufficient.
(For $H = 0$, the value of $\lambda_{\rm l}$ is irrelevant, once
${\cal L}/\lambda_{\rm s}$ has been fixed).
Thus, one needs a system of similar size as before to approach
the desired limits.

For smaller system sizes it is still possible to get accurate values of $\Delta \bar{u}/\bar{h}$ by extrapolating the computed value to the TFC limit using Eq.~(\ref{eq:tfc}) and the information from Tables~\ref{table:gapLowP}, \ref{table:gapHighP}.

\begin{table}[htb]
\begin{tabular}{c|c|c|c|c|c|c}
  $H$ & $C_{\mathrm{c}}$ &$C_{\mathrm{f}}$ & $C_{\mathrm{t}}$ & $\alpha_{\mathrm{c}}$ & $\alpha_{\mathrm{f}}$ & $\alpha_{\mathrm{t}}$ \\ \hline
  0.3 & 0.1956 & 7.4279 & 0.1076 & 1 & 0.2639 & 1 \\
  0.5 & 0.1376 & 7.0747 & 0.1107 & 1 & 0.2544 & 1 \\
  0.8 & 0.0891 & 2.5573 & 0.0353 & 1 & 0.0788 & 1 \\
\end{tabular}
\caption{Coefficients and powers in Eq.~(\ref{eq:tfc}) for $\Delta \bar{u}/\bar{h}$ at $p^* = 0.007$ obtained from the GFMD results.}
\label{table:gapLowP}
\end{table}

\begin{table}[htb]
\begin{tabular}{c|c|c|c|c|c|c}
  $H$ & $C_{\mathrm{c}}$ &$C_{\mathrm{f}}$ & $C_{\mathrm{t}}$ & $\alpha_{\mathrm{c}}$ & $\alpha_{\mathrm{f}}$ & $\alpha_{\mathrm{t}}$ \\ \hline
  0.3 & 0.1166 & 0.8058 & $6.5851\cdot10^{-4}$ & 1 & 0.3009 & 1 \\
  0.5 & 0.0515 & 1.6523 & $7.0326\cdot10^{-4}$ & 1 & 0.5072 & 1 \\
  0.8 & 0.0199 & 3.4070 & $1.3495\cdot10^{-5}$ & 1 & 0.6665 & 1 \\
\end{tabular}
\caption{Coefficients and powers in Eq.~(\ref{eq:tfc}) for $\Delta \bar{u}/\bar{h}$ at $p^* = 0.007$ obtained from the GFMD results.}
\label{table:gapHighP}
\end{table}

\subsection{Extrapolated results}

\begin{table*}[!htb]
\begin{tabular}{c|l|c|c|c|c|c|c}
  Year & Authors & $H$ values & $\kappa$ & $\kappa(H = 0.8)$ & $\epsilon_{\rm c}$ & $\epsilon_{\rm f}$ & $\epsilon_{\rm t}$ \\ \hline
  1976 & Bush, Gibson, Thomas~\cite{Bush75} & 0 $\ldots$ 1 & $\equiv \sqrt{2 \pi} \approx 2.51$ & 2.51 & 0 & $\sim 1$ & 0 \\
  2001 & Persson~\cite{Persson01} & 0 $\ldots$ 1 & $\equiv \sqrt{8/\pi} \approx 1.60$ & 1.60 & 0 & 0 & 0 \\
  2004 & Hyun, Pei, Molinari, Robbins~\cite{Hyun04} & 0.3 $\ldots$ 0.9 & 2.2 $\ldots$ 1.8 & 1.8 & 0.5 & $\approx 10^{-3}$ & 1 \\
  2007 & Campa\~{n}\'{a}, M\"{u}ser~\cite{Campana07} & 0.2 $\ldots$ 0.8 & 2.09 $\ldots$ 1.98 & 1.98 & ext. & $\approx 10^{-3}$ & 1 \\
  2012 & Putignano, Afferrante 
         {\it et al.}~\cite{Putignano12} & 0.7 $\ldots$ 1 & 2 & 2 & $\ll 1$ & $\approx 10^{-2}$ & 1 \\
  2012 & Yastrebov, Anciaux, Molinari~\cite{Yastrebov12} & 0.2 $\ldots$ 0.84 & 2.7 $\ldots$ 2.3 & $\approx 2.65$ & $\ll 1$ & 0.5 \ldots 0.01 & $\ll 1$ \\
  2013 & Current work & 0 $\ldots$ 0.8 & 2.16 $\ldots$ 1.93 & 1.93 & ext. & ext. & ext.  \\
\end{tabular}
\caption{Values of $\kappa$ at $p/E^* \approx 0.01$ obtained by different authors.
Note that for Persson theory, $\kappa$ does not depend on the choice of $\epsilon_{\rm f}$.
The notation ``$\ll 1$'' means the value of a correction which is close enough to
the corresponding limit such that the error due to the correction is less than
about 5~\%. For example, at low pressures $\epsilon_{\rm c} = 1/32$  and
$\epsilon_{\rm t} = 1/4$ correspond to ``$\ll 1$''.
The term ``ext'' means that an extrapolation to the corresponding limit is made.}
\label{table:kappa}
\end{table*}

We present reference data for the relative contact area and the mean gap. From the previous sections we conclude that extrapolating these quantities to the TFC limit does not have simple and universal rules. Therefore, to get a reasonably small error with the smallest possible computational expenditures in a contact mechanics simulation, one has to take into account the quantity of interest, the statistical properties of the surface roughness (i.e., the Hurst exponent and $\bar{g}$) and the pressure. For example, for the mean gap at low pressures and $H \lesssim 1$, one has to be careful with the thermodynamic correction, while at higher pressures and $H \gtrsim 0$ most attention should be paid to the continuum correction. We tried to satisfy the corresponding conditions for each data point in the reference plots shown below.

Figures~\ref{fig:areaVsPlow} and~\ref{fig:nonContVSinvP} show the dependence of the relative contact and non-contact area on the dimensionless pressure obtained using GFMD for systems with several different values of the Hurst exponent. Predictions of Persson theory are also presented in the same plot, for which, as was mentioned before, the contact area does not depend on the Hurst exponent $H$. This is consistent with the GFMD results, which suggest that in the TFC limit the contact area for $H$ close to 0 and to 1 should not differ by more than about 10\%.
Both our simulations and Persson theory show the contact area to be a linear function of the pressure at low loads (Fig.~\ref{fig:areaVsPlow}),
which is consistent with previous studies~\cite{Persson06,Putignano12,Yastrebov12}. At low pressures the contact area obtained using GFMD is somewhat higher than that predicted by Persson, which is consistent with the fact that according to GFMD $\kappa \approx 2$, while Persson predicts $\kappa = \sqrt{8/\pi}\approx 1.6$. Close to the complete contact at high pressures (Fig.~\ref{fig:nonContVSinvP}) the contact area changes nonlinearly with pressure. In this regime Persson theory is closer to the GFMD results, which is also consistent with the results of the previous section.

\begin{figure}[htb]
\includegraphics[width=0.46\textwidth]{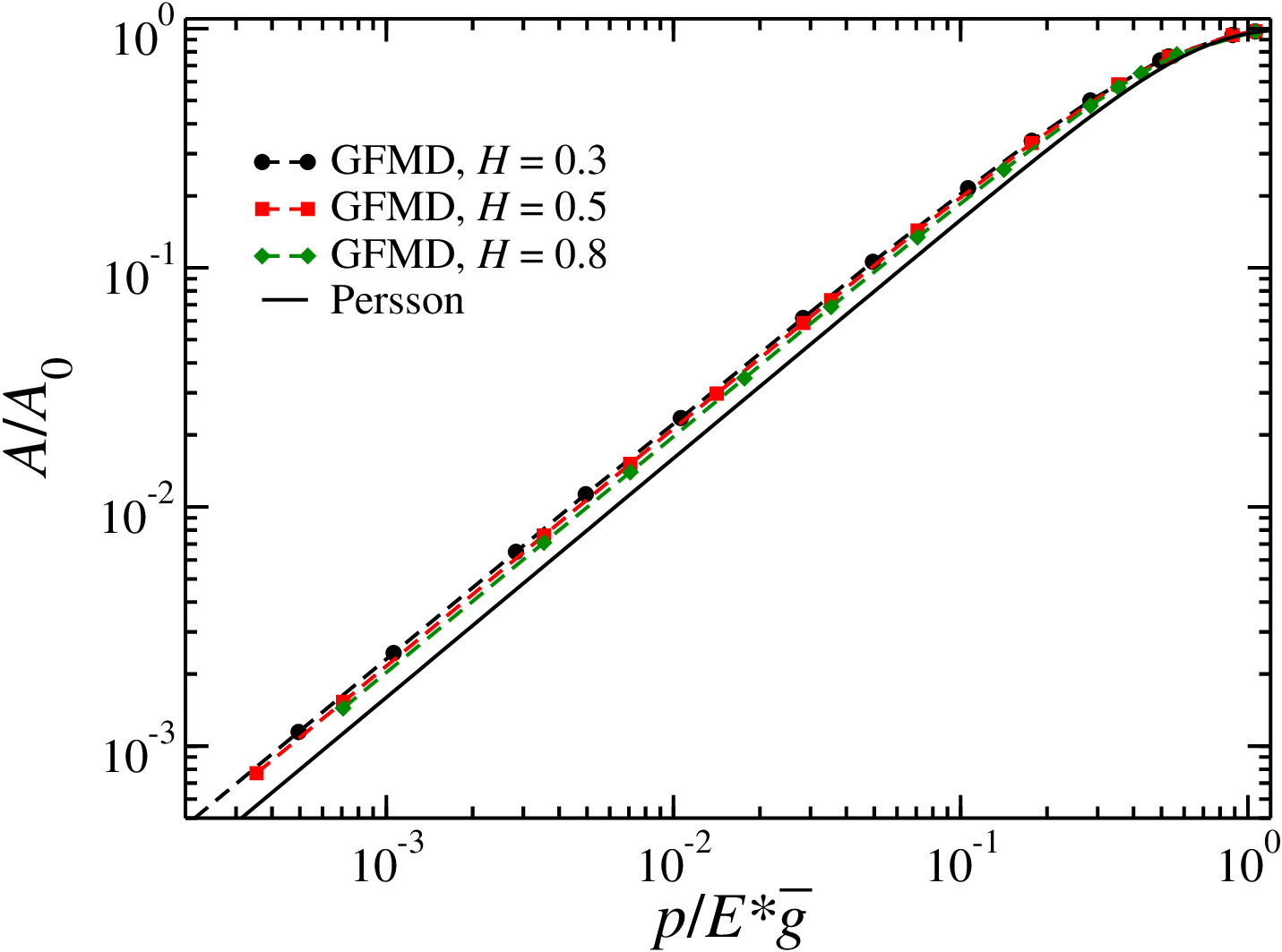}
\caption{
Relative contact area as a function of the dimensionless pressure (at low pressures) obtained using GFMD simulations as well as predicted by the Persson theory. Both theory and simulations indicate a linear dependence, for Persson theory with a slope of 1.6, while the simulations give a slope of $1.93 .. 2.16$. The dependence on $H$ is minor. High pressures are shown in Fig.~\ref{fig:nonContVSinvP}.
\label{fig:areaVsPlow}
}
\end{figure}

\begin{figure}[htb]
\includegraphics[width=0.47\textwidth]{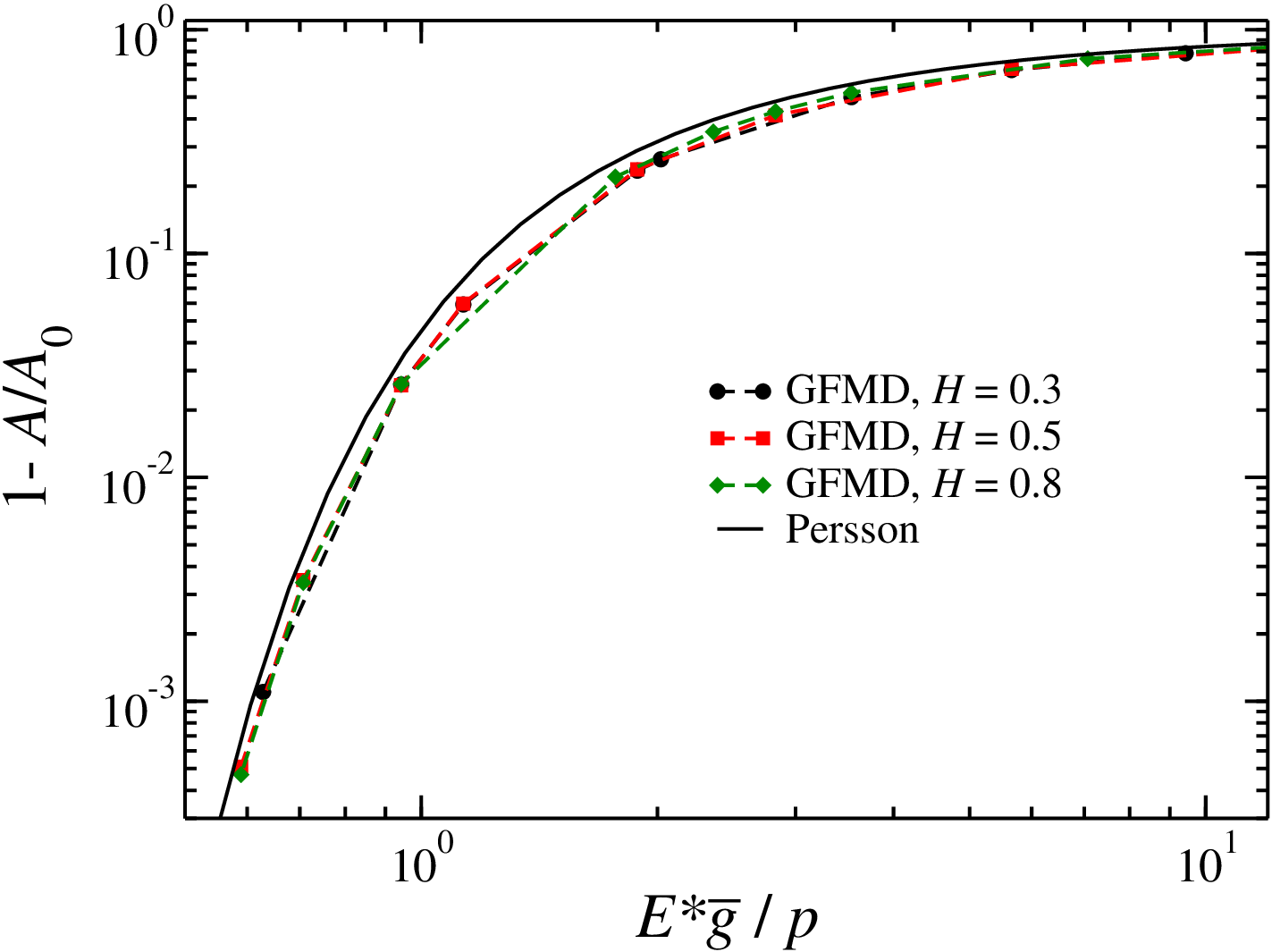}
\caption{
Continuation of Fig.~\ref{fig:areaVsPlow}. In order to better show the non-linearity both in Persson theory and simulations, we plot the \textit{non-contact} area vs. the \textit{inverse} pressure.
\label{fig:nonContVSinvP}
}
\end{figure}

A concise summary of the effects of the TFC corrections on $a_{\mathrm{r}}$ is given in Table~\ref{table:kappa}. It contains the values of $\kappa$ obtained at $p/E^* \approx 0.01$ by different authors. The data is presented in a chronological order and reflects the historical development of the insight to the problem. The corollary is that in the last few years most of the research groups have been performing computations close to the continuum limit. However, this has been often achieved by sacrificing either the fractal or the thermodynamic limit because of the scarce computational resources.
According to our results, it is more important to satisfy the fractal limit because the fractal correction $\epsilon_{\rm f}$ leads to much larger errors than the thermodynamic one,
unless pressures are extremely small.

Representing reference data for the mean gap meets some complications. As was shown in the previous section,
$\Delta \bar{u}/\bar{h}\rightarrow0$ in the fractal limit, following a power law with an exponent depending on $H$.
This means that for $\Delta \bar{u}$ it is not possible to find a normalization factor that would allow superimposing the curves for different values of $H$, and necessitates a separate plot for each value of $H$. Additionally, as real surfaces have a limited range of self-affinity (which means that the fractal limit is never reached in practice, even though $\epsilon_{\rm f}$ may be as small as $10^{-6}$) and the functional dependence of the approaching the fractal limit is also pressure-dependent, it may also be helpful to have reference data for several values of $\epsilon_{\rm f}$.

\begin{figure}[htb]
\includegraphics[width=0.45\textwidth]{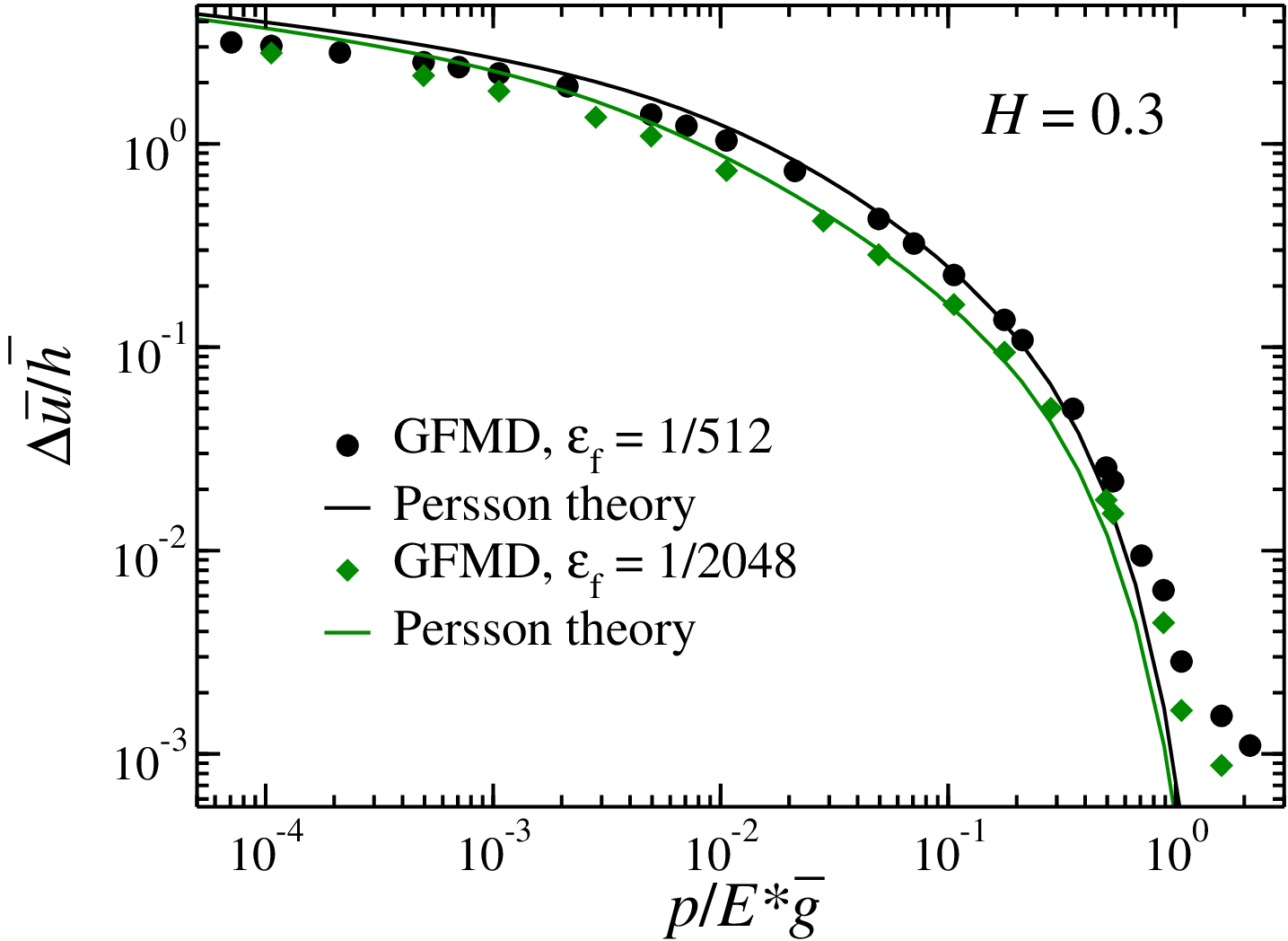}
\caption{
Pressure as a function of the mean gap at different values of the fractal correction and $H = 0.3$. The continuum and thermodynamic corrections are close to the corresponding limits.
\label{fig:gapVsPH03}
}
\end{figure}

\begin{figure}[htb]
\includegraphics[width=0.45\textwidth]{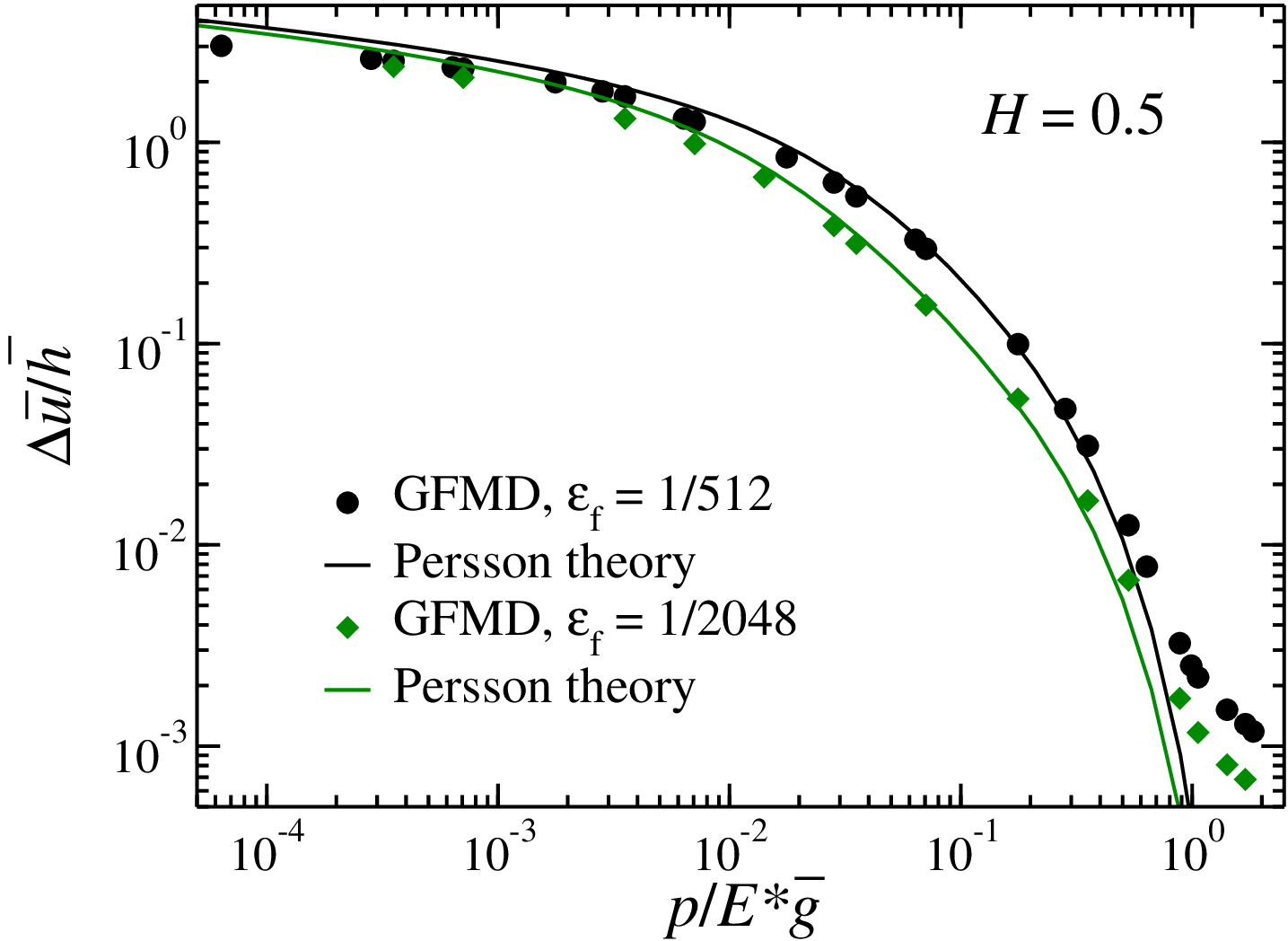}
\caption{
Pressure as a function of the mean gap at different values of the fractal correction and $H = 0.5$. The continuum and thermodynamic corrections are close to the corresponding limits.
\label{fig:gapVsPH05}
}
\end{figure}

\begin{figure}[htb]
\includegraphics[width=0.45\textwidth]{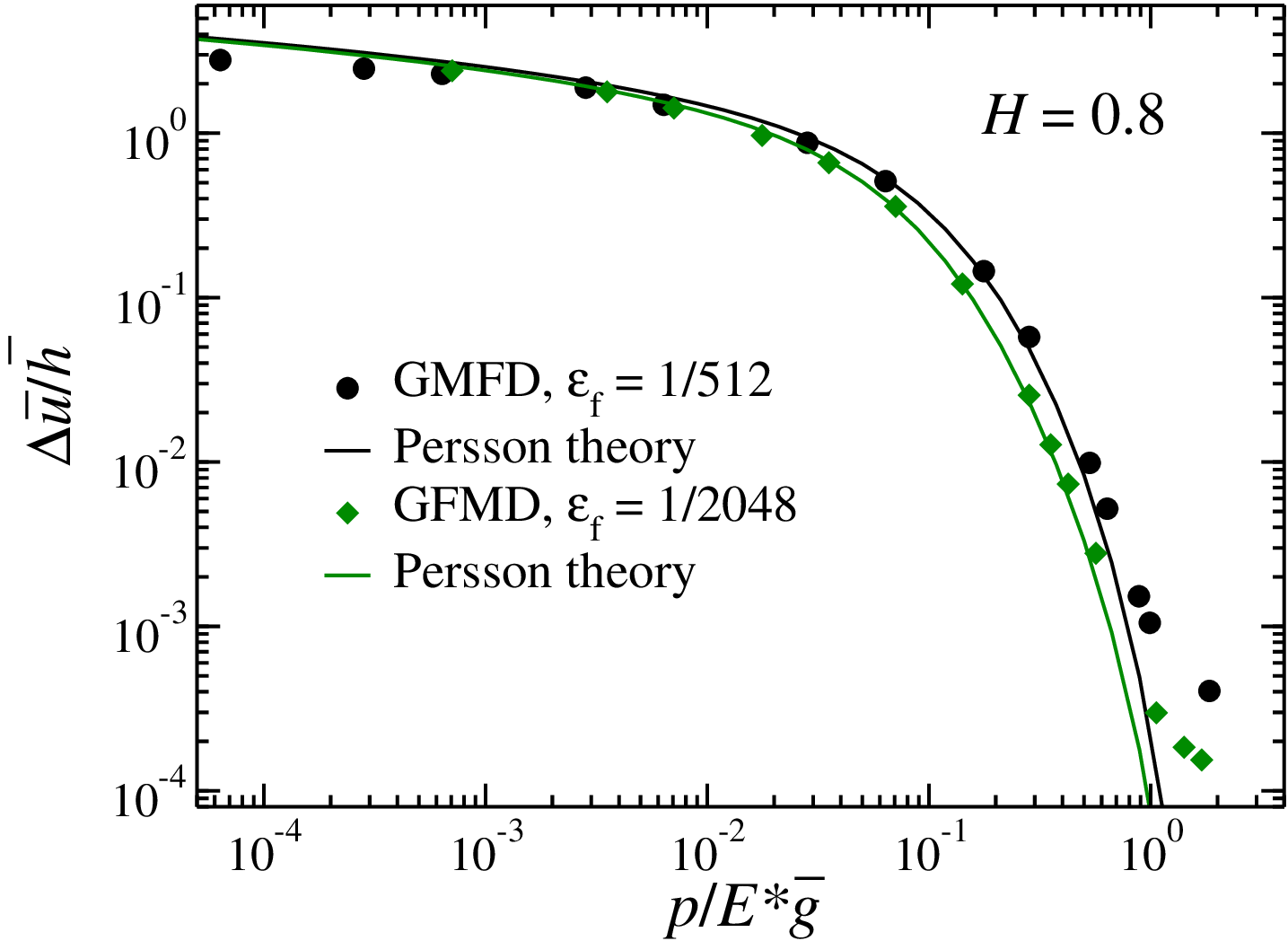}
\caption{
Pressure as a function of the mean gap at different values of the fractal correction and $H = 0.8$. The continuum and thermodynamic corrections are close to the corresponding limits.
\label{fig:gapVsPH08}
}
\end{figure}

We present such a set of reference data in Figs.~\ref{fig:gapVsPH03},~\ref{fig:gapVsPH05}, and~\ref{fig:gapVsPH08}
along with the Persson theory predictions. The $\Delta \bar{u}$ versus $p^*$ curves obtained in simulations
and theory match within about 20\% over almost the whole pressure range (except for the highest $p^*$) and have the functional form established in the
literature~\cite{Almqvist11}. Specifically, at low pressures (but high enough to avoid finite-size
effects~\cite{Pastewka13}) there is a logarithmic region, while at higher pressures a more complicated non-linear
dependence exists. As one would expect from the previous section, $\Delta \bar{u}$ decreases while approaching the
fractal limit for all the pressures. However, we did not find a simple normalization factor that depends
solely on $\epsilon_{\rm f}$ such that the curves with different $\epsilon_{\rm f}$ (but with the same $H$) would
superimpose in the whole pressure range.

We stress that all points stem from simulations that fulfill the TC limit within $10\%$.
As an example, for low pressures and $H=0.8$, the continuum limit necessitates
$\epsilon_{\rm c} \le 1/32$, and therefore very large systems, while for higher pressure
$\epsilon_{\rm c} \le 1/4$ is sufficient. In all cases $\epsilon_{\rm t} \le 1/2$ suffices.

\section{Conclusions}\label{sec:conclusions}

In this work we review analytical and computational techniques
as well as present results of GFMD calculations for two interfacial quantities
--- the relative contact area $a_{\rm r}$ and the mean gap $\Delta \bar{u}$.
The contact stiffness $K$, which is the derivative of $\Delta \bar{u}(p)$,
is implicitly given by our data.
We show that it is possible to considerably diminish the number of quantities
necessary for the description of a contact mechanics problem by choosing
proper units of measurements.
In particular, the pressure should be expressed as a multiple of the effective
elastic modulus times the RMS gradient, at least in the absence of adhesion.
The proper choice for the unit of length is less obvious.
While the RMS height is the intuitive choice, it does not turn out to
be sufficient when expressing the mean interfacial gap.
In fact we find that in the fractal
limit $\epsilon_{\rm f} \equiv \lambda_{\rm s}/\lambda_{\rm l} \to 0$,
only short-wavelength properties matter, i.e.,
$\Delta \bar{u}$ is proportional to $\bar{g}\lambda_{\rm s}$.

The observation that $\Delta \bar{u}$ is determined by short-wavelength
properties of the height spectra in the fractal limit should not let
one conclude that precise knowledge of the roughness at small length scales
is needed to predict $\Delta \bar{u}$ in practice.
Most surfaces have roughness exponents around $H = 0.8$ and most relative
contact areas for practical applications tend to be much below 1\%.
For $H = 0.8$ and $a_{\rm r} = {\cal O}(0.01)$, we find that the roughness spectrum
must be self-affine over 14 decades, e.g., from nanometer to hundreds
of kilometers in order to reach the fractal limit.
Thus, most applications should be very far from it,
unless $H$ is small or surfaces are unusually soft.
Knowledge of $\epsilon_{\rm f}$  is therefore needed to predict
$\Delta \bar{u}$ as a function of $H$ and $p/\bar{g}E^*$ in addition
to either $\bar{g}{\lambda}_{\rm s}$ or $\bar{h}$.
If $\bar{g}$ cannot  be resolved accurately down to the smallest scales,
estimates for $\Delta \bar{u}$ can still be accurate, as that quantity
is determined by roughness on mesoscopic scales.
This behavior differs from that of $a_{\rm r}$, which turns out to be inversely
proportional to $\bar{g}$ and thus to be dominated by small wavelength
properties of the surface spectra.
In the latter case fractal corrections are relatively small, even if they
are still more serious than thermodynamic and finite-size corrections.

In the present work we not only investigated the relevance of
the ratio $\lambda_{\rm s}/\lambda_{\rm l}$
but also the importance of continuum corrections
as well as finite-size or thermodynamic corrections.
The contribution of a \textit{particular} correction depends on a number
of factors, such as the pressure and the Hurst exponent and there are
no general simple rules to select the appropriate values for
$\epsilon_{\rm c}$ and $\epsilon_{\rm t}$.
However, the finite-size corrections tend to be the least problematic.
Choosing the system size twice $\lambda_{\rm l}$ is sufficient
to see the well-established linearity between normal pressure and
contact stiffness down to ${\cal O}(0.01\%)$ relative contact area.
Finite-size effects are only significant when contact is
localized near the highest asperity.
However, macroscopic surfaces must rest at least on three points
to be mechanically stable so that averaging over at least three
microscopic points of contact should be given.
For a detailed discussion on how to include finite-size effects
into contact mechanics, we point the reader to Ref.~\cite{Pastewka13}.

One purpose of this work has been to further explore the validity
of Persson theory of contact mechanics.
We find that it is not only suitable to describe the contact area
but also the non-contact area at high pressure. 
The first finding was to be expected, as Persson theory is
valid at small $p^*$ and moreover becomes exact at full contact.
However, this does not imply that the {\it deviation} from full contact
is also predicted correctly,
i.e., the asymptotic behavior of the non-contact area at high pressure.
Moreover, we observe that the dependence of $\Delta \bar{u}$
on the fractal correction -- at fixed values for $H$, $\bar{h}$, and $p^*$ --
is predicted correctly, that is, it finds the correct
functional dependence and exponents.
Other numerical parameters, such as prefactors, are only slightly off.
Merely the mean gap for relative contact areas close to unity does
not appear to match the trends conveyed by the numerical results.
This, however, only occurs for relative contact areas greater than
90\%, which is an irrelevant regime for applications.
Another point of criticism --- not further elaborated herein ---
relates to the pressure distribution.
It deviates from the predicted linear scaling at small pressure.
For brevity, we chose to not present this here but rather to 
discuss it in future work together with
an in-depth analysis of how to modify Persson theory to reflect
the observed trends.
However, concerning the observables investigated in this work, 
the current Persson theory already provides an excellent description.

\begin{acknowledgements}
We thank the J\"ulich Supercomputing Centre for computing time on JUGENE, JUQUEEN and
JUROPA. MHM also thanks DFG for support through grant No. Mu 1694/5-1.
\end{acknowledgements}

\appendix
\section{Appendix. Persson contact mechanics theory for contact area and mean gap}
\label{append:persson}
A promising approach to contact mechanics and related topics is
Persson theory~\cite{Persson01,Persson06,Almqvist11}.
The principal idea is to investigate how distributions,
such as pressure and gap distribution functions, broaden when
roughness is ignored initially but then is included gradually by
considering roughness at larger and larger wave vectors --- or greater ``magnification.''
Here, we will summarize those aspects of Persson theory
which pertain to the reference data presented in the result section, namely for contact area and mean gap. Unlike the original literature, our presentation will be focused
on the use of the dimensionless variables introduced in the main text.

Consider a contact in which the pressure distribution
is locally constant, i.e., $p(x_0\pm \lambda,y_0\pm \lambda) \approx p_0$.
One can then approximate the pressure distribution function locally
with $\Pr(\vert {\bf r} - {\bf r}_0 \vert < \lambda, p) = \delta(p-p_0)$.
Now assume that we add some roughness to the interface
by adding a Fourier component $\tilde{h}({\bf q})\exp(i{\bf q}\cdot {\bf r})$
to the roughness, where $q = 2\pi/\lambda$.
If the amplitude $\tilde{h}({\bf q})$ is small, contact in the domain
will remain essentially perfect.
This leads to a change of the local stress, see
Eqs.~(\ref{eq:elaEnerg2D}) and (\ref{eq:stress}), according to
\begin{equation}
p(\vert {\bf r} - {\bf r}_0 \vert < \lambda)
\approx p({\bf r}_0) + \frac{E^*}{2} q \tilde{h}({\bf q})
\exp(i{\bf q}\cdot {\bf r}).
\end{equation}
This means that in the vicinity of ${\bf r}_0$ there is no change
of the pressure, but there is a broadening of the pressure
distribution. In other words, the average pressure remains
$p_0$, but the second moment increases from $\Delta p_{\rm old}^2 = 0$
to
$\Delta p_{\rm new}^2 =
 \Delta p_{\rm old}^2 + \vert E^*q\tilde{h}({\bf q})/2\vert^2$.
One of the main approximations of Persson theory is that
the broadening of the pressure distribution would be similar
even if $\Delta p_{\rm old}^2$ were not zero.
The pressure distribution then broadens, whenever
we include roughness at smaller scales.
Since the broadening does not depend on pressure or location,
the total broadening, averaged over the entire contact, will then be
\begin{equation}
\Delta p^2 =
\sum_{\bf q}  \left( \frac{q E^*}{2} \right)^2 C({\bf q})
= \left( \frac{E^*}{2} \right)^2 \bar{g}^2.
\label{eq:totalBroad}
\end{equation}
In the last step, we have made use of the fact that
differentiating (heights) in real space corresponds to
multiplying with wave vectors in Fourier space.

It is known from the law of large numbers that folding
distributions functions iteratively according to
\begin{equation}
{\Pr}_{\rm new}(p) = \int \mathrm{d}p' {\Pr}_{\rm old}(p') {\rm Tr}(p'\vert p)
\end{equation}
ultimately leads to a Gaussian, where
${\rm Tr}(p'\vert p)$ is the probability that the local pressure
changes from $p'$ to $p$ after (additional) roughness is included
in the calculation.
As a first approximation, one therefore finds
$\Pr(p) \approx \exp[-(p-p_0)^2/2\Delta p^2] / \sqrt{2\pi \Delta p^2}$
for the pressure distribution.

The problem of having a single Gaussian is that negative pressures
have finite probability.
However, we know  that negative pressures
are not allowed for nonadhesive hard-wall interactions.
This problem can be solved by absorbing into noncontact any part of the
pressure distribution function that becomes negative (there, the pressure is set to zero).
If two surfaces do not touch when spatial features are resolved
down to wavelength $\lambda$, they should not come back into contact
when roughness at even smaller wavelengths is resolved.
One can implement an absorbing boundary condition, similar to the way
how mirror charges are introduced in electrostatics, by subtracting
another Gaussian from the original Gaussian:
\begin{equation}
\Pr(p>0) =
 \frac{
\exp\left\{-\frac{(p-p_0)^2}{2\Delta p^2}\right\} -
\exp\left\{-\frac{(p+p_0)^2}{2\Delta p^2}\right\}
}{\sqrt{2\pi\Delta p^2}}.
\label{eq:pressDist}
\end{equation}
The effect of the mirror Gaussian is to implement the boundary
condition, while leaving the mean pressure $\bar{p} = \int p\,\mathrm{d}p \Pr(p>0)$
invariant, i.e., independent of $\Delta p$.
Persson finds Eq.~(\ref{eq:pressDist}) through a small detour by mapping the
integral equation for the broadening of the pressure distribution to a
differential equation, which is isomorphic to the diffusion equation.
This detour, however, can be avoided without loss of information.

Eq.~(\ref{eq:pressDist}) enables one to deduce the relative contact area
\begin{eqnarray}
a_{\mathrm{r}} & = & \int_{0^+}^\infty  \mathrm{d}p \Pr(p)
\end{eqnarray}
because any finite local pressure is interpreted as
occurring where the solids are in contact.
The solution  of the integral reads
\begin{eqnarray}
a_{\mathrm{r}} & = & {\rm erf}\left(\frac{p_0}{\sqrt{2}\Delta p}\right)
     =  {\rm erf}\left(\frac{\sqrt{2}p_0}{E^*\bar{g}}\right)
          \label{eq:contAreaExact} \\
    & = & \sqrt{\frac{8}{\pi}} \frac{p_0}{E^* \bar{g}} + {\cal O}\left\{
\left( \frac{p_0}{E^* \bar{g}} \right)^3\right\}.
          \label{eq:contAreaApprox}
\end{eqnarray}
Three properties of the solution are interesting to observe:
(i) it satisfies the finding of Sect.~\ref{sec:dimAnal}
that observables should depend on external pressure divided by
the product of $E^*$ and $\bar{g}$ but they should not depend
on any other dimensionless variable other than possibly $H$.
(ii) The solution turns out to not depend on $H$ for any value of $p_0$.
(iii) Corrections to the linear relationship between contact area
and pressure are only of order $p_0^3$.
This implies that linearity between load and contact area should
persist up to at least 10\% contact.
These predictions are confirmed by Fig.~\ref{fig:areaVsPlow}.

Next, we wish to express the gap as a function of normal pressure.
For $p_0 \to \infty$, the mean gap tends to zero, while
the two surfaces touch in just one point for $p_0 = 0^+$.
Given the nature of our problem, there is a monotonic dependence
of the gap on load in between the two limiting cases of no contact and full contact.
This allows us to express the work done by the pressure on the
elastic manifold as follows:
\begin{eqnarray}
\frac{1}{A_0} \mathrm{d}U_{\rm el} & = & -p(\bar{u}) \mathrm{d}\bar{u}
    = -p \frac{\mathrm{d}\bar{u}}{\mathrm{d}p} \mathrm{d}p,
\end{eqnarray}
where  $\bar{u} = u_0 + \tilde{u}({\bf q}=0)$ denotes the displacement
with respect to some well-chosen reference point $u_0$. We choose $u_0$ in such a way that it corresponds to the full contact at the external pressure $p_{\rm ref} = \infty$.
Thus, if we knew $U_{\rm el}$ as a function of $p$, we could obtain the displacement $\bar{u}$ via
\begin{equation}
\Delta \bar{u} = \bar{u}(p_0) - \bar{u}(p_{\rm ref}) =
\frac{1}{A_0}
\int_{p_0}^{p_{\rm ref}} \mathrm{d}p'_0
\frac{1}{p'_0}
\frac{\mathrm{d}U_{\rm el}}{\mathrm{d}p'_0}.
\label{eq:gap}
\end{equation}
To solve for the elastic energy, Persson argues~\cite{Persson01} that the
displacement field $\tilde{u}({\bf q})$ follows $\tilde{h}({\bf q})$
for the fraction of the interface that is in contact at
a resolution of ${\bf q}$.
Thus,
\begin{equation}
\frac{1}{A_0}U_{\rm el}\left(p_0\right) = \frac{E^*}{4}
\sum_{\bf q} \gamma(p_0,{\bf q}) \, q \, \vert\tilde{h} ({\bf q})\vert^2,
\label{eq:elasticEnergy}
\end{equation}
where
\begin{equation}
\gamma(p_0,{\bf q})=a_{\mathrm{r}}(p_0,{\bf q})[\gamma + (1-\gamma)a^2_{\mathrm{r}}(p_0,{\bf q})]
\label{eq:gamma}
\end{equation}
is an \textit{ad hoc} helper function which approximates the dependence of the elastic energy on the
resolution-dependent contact area $a_{\mathrm{r}}(p_0,{\bf q})$ in different pressure
regimes. At low external pressures $\gamma(p,{\bf q})$ is proportional to
$a_{\mathrm{r}}(p_0,{\bf q})$ while at the complete contact it is equal to 1.
The contact area $a_{\mathrm{r}}(p_0,{\bf q})$ follows from Eq.~(\ref{eq:contAreaExact})
or (\ref{eq:contAreaApprox}) by confining the evaluation of $\bar{g}$ to wave
numbers less than $q_{\mathrm{s}}$. An empirical correction factor $\gamma$ has the value of the order of unity
(a value of 0.42 has been used in the literature~\cite{Persson07PRL,Yang08JPCM})
and reflects the fact that the elastic energy stored in the contact region is less
than the average elastic energy for perfect contact~\cite{Persson07PRL,Yang08JPCM}.

Substituting Eq.~(\ref{eq:elasticEnergy}) into Eq.~(\ref{eq:gap}) and assuming an \textit{ideal self-affine surface} characterized with the power spectrum from the Eq.~(\ref{eq:scalCinRec}), after some algebra we obtain the following expression for the mean gap $\Delta \bar{u}$ as a function of the external pressure $p_0$:
\begin{align}
\Delta \bar{u}
= &\frac{\overline{h}}{\sqrt{2\pi}}\sqrt{\frac{H}{1-H}}
   \sqrt{\frac{1}{1-\epsilon_{\mathrm{f}}^{2H}}} \; \times \notag\\
  &\int_{0}^{c}\mathrm{d}k\,
   \left(k^{2}+1\right)^{1/(2H-2)}\, 
   \left[
   \gamma\,\mathrm{E}_{1}\left(x^{2}\right)\,+
   3\,(1-\gamma) \times \phantom{\int_{1}^{\infty}}\right. \notag\\
  &\left.\int_{1}^{\infty}\,
   \left[\mathrm{erf}\left(tx\right)\right]^{2}\,
   \exp\left(-t^2x^2\right)\,\frac{\mathrm{d}t}{t}
   \right],
\label{eq:gapFinal}
\end{align}
where we have replaced the Fourier sum with a Fourier integral.
Other quantities in Eq.(\ref{eq:gapFinal}) are $x \equiv \sqrt{2}\,p_{0}c/(E^*\bar{g}k)$,
$c^{2} \equiv (1-\epsilon_{\mathrm{f}}^{2-2H})/\epsilon_{\mathrm{f}}^{2-2H}$, and $\mathrm{E}_{1}$ is a variant of the exponential integral given by
\begin{align}
\mathrm{E}_{\alpha}\left(x\right) = \int_{1}^{\infty}\frac{\exp\left[-tx\right]}{t^{\alpha}}\,\mathrm{d}t,
\end{align}
and available as a special function in standard libraries (e.g., the Boost library for C++~\cite{Boost}).

Eq.~(\ref{eq:gapFinal}) is not analytically tractable. However, in future work we will investigate $\gamma(p_0,{\bf q})$
in more detail and an expression might be found that allows to simplify the integral further.

The pseudocode for computing Persson theory is as follows:
\begin{itemize}
\item Specify the input parameters\\
These mainly include the characteristics of the rough surface.
\item Express the input in the dimensionless form. 
\item Calculate the contact area using Eq.~(\ref{eq:contAreaExact})
\item Calculate the mean gap using Eq.~(\ref{eq:gapFinal})\\
Note that choosing a logarithmic mesh for the nested integral to $\infty$ in Eq.~(\ref{eq:gapFinal})
will improve efficiency without significant loss of accuracy. This effect can also be achieved through a
variable substitution of the form $p = p_0 \exp(\mu)$.
\end{itemize}

\section{Appendix. A review of Green's function molecular dynamics}
\label{sec:GFMD}

Green's function molecular dynamics (GFMD) makes it possible to calculate
the response of a semi-infinite elastic solid to external forces
acting solely on the surface~\cite{Campana06,Kong09}.
It can be described as a classical boundary value method which is
solved with regular molecular dynamics.
In principle, it is possible to simulate natural dynamics.
However, in this work we are only concerned with the static limiting
case, which is why we content ourselves with damped dynamics.
These should be set up in such a way that the static solution is found
in the quickest possible way.
Using natural dynamics would not be efficient, as these would suffer
from critical slowing down.
The number of steps to reach equilibrium would scale with the
square of the linear dimension.

In principle, GFMD attempts the solution of Eq.~(\ref{eq:stress}).
However, the interfacial pressure is not known explicitly but only
implicitly through the boundary condition Eq.~(\ref{eq:boundaryCond}).
This implies that the curvature of the potential diverges when the
two surfaces start to overlap, which, in principle, makes the use
of an infinitely small time step necessary.
In the early days of molecular and computational fluid dynamics,
several strategies were designed for related problems in
the context of hard-disk interactions~\cite{Allen87}.
One approach was to assign a coefficient of restitution to
a collision of two hard disks which specifies how much of the kinetic
energy is conserved during a collision.
For our contact mechanics problem, we set this coefficient to zero.

In the following, we will describe how we implement these dynamics
and also describe all other main aspects of our GFMD program
in terms of pseudocode.
\begin{itemize}
\item Setup of rigid substrate \vspace*{1mm} \\
Assign uniform random numbers of zero mean and finite variance
for the real and imaginary  parts of $\tilde{h}({\bf q})$.
All $\tilde{h}({\bf q})$ are divided by $q^{1+H}$.
Next, heights are transformed into real space.
For this purpose, we use the FFTW library~\cite{Frigo05fftw}.
We shift the elastic surface such that it touches the rigid
substrate in one (or more) points without applying pressure
(i.e., $h_{\rm min} = 0$). The largest height is then stored in
$h_{\rm max}$.
RMS gradient and RMS height are best evaluated
in Fourier space.
\item Setup of elastic top solid\vspace*{1mm} \\
Set all grid points to $h_{\rm max}$ and define this as a reference.
Initialize the damping $\eta$ such that the slowest mode,
i.e., the center-of-mass mode is critically damped or slightly underdamped.
This can be achieved with $\eta \propto p/E^*\bar{g}{\cal L}$,
unless $a_{\rm r}$ is close to 0. When $a_{\rm r}$ is close to 0 we use $\eta \propto \left(p/E^*\bar{g}\right)^\alpha\sqrt{\beta/{\cal L}}$, where $\alpha$ and $\beta$ are the parameters which depend on $\cal L$ and typically are found empirically.
Note that neither the mass nor the damping should be made a function
of wave vector if nonholonomic boundary conditions are in place.
\item
Loop over time steps until converged
\begin{itemize}
\item
Transform displacements into Fourier space
\item Calculate elastic restoring forces\\
$\tilde{F}({\bf q}) = - q (E^*/2) \tilde{u}_{\rm now}({\bf q})$
\item Add external pressure\\
$\tilde{F}(0) \leftarrow \tilde{F}(0) + p$
\item Add damping forces\\
$\tilde{F}({\bf q})\leftarrow \tilde{F}({\bf q}) + \eta\{\tilde{u}_{\rm now}({\bf q})
- \tilde{u}_{\rm old}({\bf q}) \}$
\item Use Verlet to solve equation of motion\\
$\tilde{u}_{\rm new}({\bf q}) =
2\tilde{u}_{\rm now}({\bf q}) - \tilde{u}_{\rm old}({\bf q})
+ \tilde{F}({\bf q}) \Delta t^2
$
\item Transform displacement into real space
\item Implement the boundary condition\\
$ u_{\rm new}({\bf r}) \leftarrow \max\{ u_{\rm new}({\bf r}) , -h({\bf r})\}$
\item Assign
$ u_{\rm old}({\bf r}) \leftarrow u_{\rm now}({\bf r})$ \\
$ u_{\rm now}({\bf r}) \leftarrow u_{\rm new}({\bf r})$
\item Check termination conditions, e.g. that\\
the contact area, defined as the relative number of points satisfying
$u({\bf r}) = -h({\bf r})$, may not have changed in many steps;
kinetic energy of each individual mode less than a threshold, etc.
\end{itemize}
\end{itemize}

\bibliographystyle{spmpsci}       


\begin{thebibliography}{10}
\providecommand{\url}[1]{#1}
\providecommand{\urlprefix}{URL }
\expandafter\ifx\csname urlstyle\endcsname\relax
  \providecommand{\doi}[1]{DOI~\discretionary{}{}{}#1}\else
  \providecommand{\doi}{DOI~\discretionary{}{}{}\begingroup
  \urlstyle{rm}\Url}\fi

\bibitem{Boost}
Boost {C}++ {L}ibraries.
\newblock \url{http://boost.org}

\bibitem{Aifantis87}
Aifantis, E.C.: The physics of plastic deformation.
\newblock International Journal of Plasticity \textbf{3}, 211 (1987)

\bibitem{Allen87}
Allen, M.P., Tildesley, D.J.: Computer Simulation of Liquids.
\newblock Oxford University Press, Oxford (1987)

\bibitem{Almqvist11}
Almqvist, A., Campa{\~n}\'a, C., Prodanov, N., Persson, B.N.J.: Interfacial
  separation between elastic solids with randomly rough surfaces: Comparison
  between theory and numerical techniques.
\newblock J. Mech. Phys. Solids \textbf{59}, 2355 (2011)

\bibitem{Barber13}
Barber, J.R.: Incremental stiffness and electrical contact conductance in the
  contact of rough finite bodies.
\newblock Phys. Rev. E. \textbf{87}, 013,203 (2013)

\bibitem{Bush75}
Bush, A.W., Gibson, R.D., Thomas, T.R.: Elastic contact of a rough surface.
\newblock Wear \textbf{35}, 87 (1975)

\bibitem{Campana06}
Campa{\~n}\'a, C., M\"user, M.H.: Practical green's function approach to the
  simulation of elastic semi-infinite solids.
\newblock Phys. Rev. B \textbf{74}, 075,420 (2006)

\bibitem{Campana07}
Campa{\~n}\'a, C., M\"user, M.H.: Contact mechanics of real vs. randomly rough
  surfaces: A {G}reen's function molecular dynamics study.
\newblock Europhys. Lett. \textbf{77}, 38,005 (2007)

\bibitem{Campana11}
Campa{\~n}\'a, C., Persson, B.N.J., M\"user, M.H.: Transverse and normal
  interfacial stiffness of solids with randomly rough surfaces.
\newblock J. Phys.: Condens. Matter \textbf{23}, 085,001 (2011)

\bibitem{Dapp12PRL}
Dapp, W.B., L\"ucke, A., Persson, B.N.J., M\"user, M.H.: Self-affine elastic
  contacts: percolation and leakage.
\newblock Phys. Rev. Lett. \textbf{108}, 244,301 (2012)

\bibitem{Frigo05fftw}
Frigo, M., Johnson, S.G.: The design and implementation of fftw3.
\newblock Proceedings of the IEEE \textbf{93}(2), 216 (2005)

\bibitem{Greenwood66}
Greenwood, J.A., Williamson, J.B.P.: Contact of nominally flat surfaces.
\newblock Proc. R. Soc. London \textbf{A295}, 300 (1966)

\bibitem{Hyun04}
Hyun, S., Pei, L., Molinari, J.F., Robbins, M.O.: Finite-element analysis of
  contact between elastic self-affine surfaces.
\newblock Phys. Rev. E \textbf{70}, 026,117 (2004)

\bibitem{Johnson85}
Johnson, K.L.: Contact Mechanics.
\newblock Cambridge University Press, New York (1985)

\bibitem{Kendall01}
Kendall, K.: Molecular Adhesion and its Applications: The Sticky Universe.
\newblock Kluwer Academic, New York (2001)

\bibitem{Kong09}
Kong, L.T., Bartels, G., Campa{\~n}\'a, C., Denniston, C., M\"user, M.H.:
  Implementation of green's function molecular dynamics: An extension to
  lammps.
\newblock Computer Physics Communications \textbf{180}, 1004 (2009)

\bibitem{Landau70v7}
Landau, L.D., Lifshitz, E.M.: Theory of Elasticity, 3rd ed.
\newblock Pergamon Press, Oxford (1970)

\bibitem{Lechenault10}
Lechenault, F., Pallares, G., George, M., Rountree, C., Bouchaud, E., Ciccotti,
  M.: Effects of finite probe size on self-affine roughness measurements.
\newblock Phys. Rev. Lett. \textbf{104}, 025,502 (2010)

\bibitem{Lorenz13jpcm}
Lorenz, B., Krick, B.A., Mulakaluri, N., Smolyakova, M., Dieluweit, S., Sawyer,
  W.G., Persson, B.N.J.: Adhesion: role of bulk viscoelasticity and surface
  roughness.
\newblock J. Phys.: Condens. Matter \textbf{25}, 225,004 (2013)

\bibitem{Lorenz10EPJE}
Lorenz, B., Persson, B.N.J.: Leak rate of seals: Effective-medium theory and
  comparison with experiment.
\newblock Eur. Phys. J. E \textbf{31}, 159 (2010)

\bibitem{Lorenz10EPJE2}
Lorenz, B., Persson, B.N.J.: Time-dependent fluid squeeze-out between solids
  with rough surfaces.
\newblock Eur. Phys. J. E \textbf{32}, 281 (2010)

\bibitem{Lyashenko13}
Lyashenko, I., Pastewka, L., Persson, B.N.J.: On the validity of the method of
  reduction of dimensionality: Area of contact, average interfacial separation
  and contact stiffness.
\newblock Tribology Letters \textbf{52}, 223--229 (2013)

\bibitem{Ma12}
Ma, Z.S., Zhou, Y.C., Long, S., Lu, C.: On the intrinsic hardness of a metallic
  film/substrate system: Indentation size and substrate effects.
\newblock International Journal of Plasticity \textbf{34}, 1 (2012)

\bibitem{Pastewka13}
Pastewka, L., Prodanov, N., Lorenz, B., M\"user, M.H., Robbins, M.O., Persson,
  B.N.J.: Finite-size scaling in the interfacial stiffness of rough elastic
  contacts.
\newblock Phys. Rev. E \textbf{87}, 062,809 (2013)

\bibitem{Persson01}
Persson, B.N.J.: Theory of rubber friction and contact mechanics.
\newblock J. Chem. Phys. \textbf{115}, 3840 (2001)

\bibitem{Persson06}
Persson, B.N.J.: {Contact mechanics for randomly rough surfaces}.
\newblock Surf. Sci. Rep. \textbf{61}, 201 (2006)

\bibitem{Persson07PRL}
Persson, B.N.J.: Relation between interfacial separation and load: A general
  theory of contact mechanics.
\newblock Phys. Rev. Lett. \textbf{99}, 125,502 (2007)

\bibitem{Persson05JPCM}
Persson, B.N.J., Albohr, O., Tartaglino, U., Volokitin, A.I., Tosatti, E.: On
  the nature of surface roughness with application to contact mechanics,
  sealing, rubber friction and adhesion.
\newblock J. Phys. Condens. Matter \textbf{17}, R1--R62 (2005)

\bibitem{Persson12EPJE}
Persson, B.N.J., Prodanov, N., Krick, B.A., Rodriguez, N., Mulakaluri, N.,
  Sawyer, W.G., Mangiagalli, P.: Elastic contact mechanics: Percolation of the
  contact area and fluid squeeze-out.
\newblock Eur. Phys. J. E \textbf{35}, 5 (2012)

\bibitem{Persson09JPCM}
Persson, B.N.J., Scaraggi, M.: On the transition from boundary lubrication to
  hydrodynamic lubrication in soft contacts.
\newblock J. Phys. Condens. Matter \textbf{21}, 185,002 (2009)

\bibitem{Persson01adhes}
Persson, B.N.J., Tosatti, E.: The effect of surface roughness on the adhesion
  of elastic solids.
\newblock J. Chem. Phys. \textbf{115}(12), 5597 (2001)

\bibitem{Persson08JPCM}
Persson, B.N.J., Yang, C.: Theory of the leak-rate of seals.
\newblock J. Phys. Condens. Matter \textbf{20}, 315,011 (2008)

\bibitem{Power91}
Power, W.L., Tullis, T.E.: Euclidean and fractal models for the description of
  rock surface roughness.
\newblock Journal of Geophysical Research \textbf{96}, 415 (1991)

\bibitem{Prodanov13}
Prodanov, N., Gachot, C., Rosenkranz, A., M\"ucklich, F., M\"user, M.H.:
  Contact mechanics of laser-textured surfaces: Correlating contact area and
  friction.
\newblock Tribol. Lett. \textbf{50}, 41 (2013)

\bibitem{Putignano12}
Putignano, C., Afferrante, L., Carbone, G., Demelio, G.: The influence of the
  statistical properties of self-affine surfaces in elastic contacts: A
  numerical investigation.
\newblock J. Mech. Phys. Solids \textbf{60}, 973 (2012)

\bibitem{Putignano13}
Putignano, C., Afferrante, L., Carbone, G., Demelio, G.: A multiscale analysis
  of elastic contacts and percolation threshold for numerically generated and
  real rough surfaces.
\newblock Tribology International \textbf{64}, 148 (2013)

\bibitem{Yang08JPCM}
Yang, C., Persson, B.N.J.: Contact mechanics: contact area and interfacial
  separation from small contact to full contact.
\newblock J. Phys. Condens. Matter \textbf{20}, 215,214 (2008)

\bibitem{Yastrebov12}
Yastrebov, V.A., Anciaux, G., Molinari, J.F.: Contact between representative
  rough surfaces.
\newblock Phys. Rev. E \textbf{86}, 035,601R (2012)

\end{thebibliography}
\end{document}